\documentclass[prd,preprint,amsmath,amssymb,floatfix,nopreprintnumbers,showpacs]{revtex4}
\usepackage{graphicx}
\usepackage{longtable}

\newcommand{\beq}{\begin{equation}}
\newcommand{\eeq}{\end{equation}}
\newcommand{\beqs}{\begin{eqnarray}}
\newcommand{\eeqs}{\end{eqnarray}}

\newcommand{\errpm}[2]{$\left(\begin{smallmatrix}#1\\#2\end{smallmatrix}\right)$}

\newcommand{\chidof}{\ensuremath{\mbox{$\chi^2/\text{DOF}$}}}

\begin{document}

\title{Lattice study of conformal behavior in SU$(3)$ Yang-Mills theories}

\author{Thomas Appelquist}
%\email{Thomas.Appelquist@Yale.edu}
\affiliation{Department of Physics, Sloane Laboratory, Yale University,
           New Haven, Connecticut, 06520, USA}

\author{George T.\ Fleming}
%\email{George.Fleming@Yale.edu}
\affiliation{Department of Physics, Sloane Laboratory, Yale University,
           New Haven, Connecticut, 06520, USA}

\author{Ethan T.\ Neil}
%\email{Ethan.Neil@Yale.edu}
\affiliation{Department of Physics, Sloane Laboratory, Yale University,
           New Haven, Connecticut, 06520, USA}

%%%%%%%%%%%%%%%%%%%%%%%%%%%%%%%%%%%%%%%%%%%%%%%%%%%%%%%%%%%%%%%%%%%%%%%%%%%%%%
\begin{abstract}
%%%%%%%%%%%%%%%%%%%%%%%%%%%%%%%%%%%%%%%%%%%%%%%%%%%%%%%%%%%%%%%%%%%%%%%%%%%%%%

Using lattice simulations, we study the extent of the conformal window for an
$\text{SU}(3)$ gauge theory with $N_f$ Dirac fermions in the fundamental
representation. We extend our recently reported work, describing the general
framework and the lattice simulations in more detail. We find that the
theory is conformal in the infrared for $N_f = 12$, governed by an infrared fixed point, whereas
the $N_f = 8$ theory exhibits confinement and chiral symmetry breaking.
We therefore conclude that the low end of the conformal window $N_f^\text{c}$ lies in the range $8 \leq N_f^\text{c}
\leq 12$. We discuss open questions and the potential relevance of the present
work to physics beyond the standard model.

%%%%%%%%%%%%%%%%%%%%%%%%%%%%%%%%%%%%%%%%%%%%%%%%%%%%%%%%%%%%%%%%%%%%%%%%%%%%%%
\end{abstract}
%%%%%%%%%%%%%%%%%%%%%%%%%%%%%%%%%%%%%%%%%%%%%%%%%%%%%%%%%%%%%%%%%%%%%%%%%%%%%%

%%% GTF: choose correct order for PACS

\maketitle

%%%%%%%%%%%%%%%%%%%%%%%%%%%%%%%%%%%%%%%%%%%%%%%%%%%%%%%%%%%%%%%%%%%%%%%%%%%%%%
\section{Introduction}\label{sec:intro}
%%%%%%%%%%%%%%%%%%%%%%%%%%%%%%%%%%%%%%%%%%%%%%%%%%%%%%%%%%%%%%%%%%%%%%%%%%%%%%

The conformal window in a gauge field theory with $N_f$ light fermions is the
range of $N_f$ values such that the theory is asymptotically free and the
infrared coupling is governed by an infrared fixed point. In an $\text{SU}(N)$
gauge theory with $N_f$ Dirac fermions in the fundamental representation, the
conformal window extends from $11N/2$ down to some critical value $N_f^\text{c}$ at
which a transition is expected to a phase in which chiral symmetry is broken
spontaneously, and confinement sets in. In a recent paper \cite{Appelquist:2007hu},
we provided the first nonperturbative evidence, using lattice simulations, that the
lower end of the conformal window for the $\text{SU}(3)$ gauge theory lies in the
range $8 < N_f^\text{c} < 12$.

Gauge theories in or near the conformal window could play a key role in
describing physics beyond the standard model. If the fermions carry
electroweak quantum numbers, and if $N_f$ lies outside but near the conformal
window, then the theory could drive electroweak breaking, forming the basis of
walking technicolor theories. If the fermions do not carry electroweak quantum
numbers, then $N_f$ could lie within the conformal window, and the theory
could describe some new, conformal sector, possibly coupled to the standard
model through $\text{SU}(N)$ invariant operators.  It is important to learn as much
as possible about the extent of the conformal window in these theories, as
well as the order of the transition at $N_f^\text{c}$ and the properties of the
theory within the window and near it.

To obtain the result $8 < N_f^\text{c} < 12$ for Dirac fermions in the fundamental
representation of an $\text{SU}(3)$ gauge group, we employed \cite{Appelquist:2007hu}
a gauge invariant, nonperturbative running coupling derived from the
Schr\"odinger functional of the gauge theory
\cite{Luscher:1992an,Sint:1993un,Bode:2001jv}. Defined within a Euclidean box
of volume $O(L^4)$, it avoids typical finite volume effects by using $L$
itself as the sliding scale. For the asymptotically free theories being
considered, it agrees with the perturbative running coupling coupling at small
enough $L$, and can be used to probe for conformal behavior in the large $L$ limit. We made use of staggered fermions as in Ref.~\cite{Heller:1997pn}, and
therefore restricted attention to values of $N_f$ that are multiples of $4$.
The value $N_f = 16$ leads to an infrared fixed point that is so weak that it
is best studied in perturbation theory.  The value $N_f = 4$ is expected to be
well outside the conformal window, leading to confinement and chiral symmetry
breaking \cite{Sui:2001rf} as with $N_f = 2$.  We thus focused on the two
values $N_f = 8$ and $N_f = 12$. We argued \cite{Appelquist:2007hu} that for
$N_f = 12$, the theory is indeed conformal in the infrared.  For $N_f = 8$, we showed, in
disagreement with an earlier lattice study \cite{Iwasaki:2003de}, that the
theory breaks chiral symmetry and confines. There is no evidence for an
infrared fixed point.

In this paper, we provide a more detailed description of the results of Ref.
\cite{Appelquist:2007hu}, and extend the analysis in several ways.  Continuing
to focus on the values $N_f = 8$ and $12$, we describe more extensive
numerical simulations of the running coupling and discuss in more detail the
treatment of lattice artifacts and the extrapolation to the continuum. We
again conclude that $8 < N_f^\text{c} < 12$, with a more precise determination of
the large $L$ behavior of the running coupling.

This work paves the way for future $\text{SU}(3)$ simulations at other values of
$N_f$, in particular, in the range between $8$ and $12$, for fermions in other
representations of the $\text{SU}(3)$ gauge group, and for other gauge groups.
Simulations using other definitions of the running coupling, for example,
derived from the Wilson loop, should also be carried out
\cite{Bilgici:2008mt}.  Conclusions about the conformal window based on the
study of running couplings should be confirmed by zero-temperature lattice
simulations of the chiral condensate and other quantities. These include the
particle masses and Goldstone boson decay constants for $N_f$ just below
$N_f^\text{c}$, and various correlation functions within the conformal window
\cite{Fleming:2008jx}.

In Sec.~\ref{sec:pt}, we describe what is known from perturbative and other studies
about the conformal window in $\text{SU}(N)$ gauge theories. For comparison, we
describe briefly the conformal window in supersymmetric $\text{SU}(N)$ gauge
theories. In Sec.~\ref{sec:sf}, we review the Schr\"odinger functional framework
\cite{Luscher:1992an,Sint:1993un,Bode:2001jv} for our numerical simulations.
Our lattice simulations are described in Sec.~\ref{sec:lat}. We discuss the
algorithms, the use of staggered fermions, the continuum extrapolation, and analysis methods.
In Sec.~\ref{sec:res}, we present our results for both $N_f = 8$ and $12$, and compare to other studies. We
summarize our work and discuss open questions in Sec.~\ref{sec:summ}.  The details of our
error analysis are given in Appendix~\ref{app:error}, and tables of simulation data appear
in Appendix~\ref{app:data}.

%%%%%%%%%%%%%%%%%%%%%%%%%%%%%%%%%%%%%%%%%%%%%%%%%%%%%%%%%%%%%%%%%%%%%%%%%%%%%%
\section{The Conformal Window}\label{sec:pt}
%%%%%%%%%%%%%%%%%%%%%%%%%%%%%%%%%%%%%%%%%%%%%%%%%%%%%%%%%%%%%%%%%%%%%%%%%%%%%%

We first review what is known from the perturbative
expansion of the beta function, and then discuss briefly a nonperturbative
approach based on the counting of degrees of freedom and some other,
quasiperturbative studies.

%%%%%%%%%%%%%%%%%%%%%%%%%%%%%%%%%%%%%%%%%%%%%%%%%%%%%%%%%%%%%%%%%%%%%%%%%%%%%%
\subsection{Perturbation theory}
%%%%%%%%%%%%%%%%%%%%%%%%%%%%%%%%%%%%%%%%%%%%%%%%%%%%%%%%%%%%%%%%%%%%%%%%%%%%%%

The existence of a conformal window in $\text{SU}(N)$ gauge theories has been known
since the computation of the two-loop beta function by Caswell in 1974
\cite{Caswell:1974gg}. If the number of massless fermions $N_f$ is near but
just below the number $N_f^\text{af}$ at which asymptotic freedom sets in, then the
two-loop term is opposite in sign to the one-loop term and the resultant infrared
fixed point is weak, accessible in perturbation theory. There is no
confinement and chiral symmetry is unbroken. The properties of this phase were
studied by expanding in $N_f^\text{af} - N_f$ in Ref. \cite{Banks:1981nn}.

As $N_f$ is reduced, the strength of the infrared fixed point grows, with $N_f$
ultimately reaching the value $N_f^\text{c}$ at which the transition to the chirally
broken and confining phase is thought to set in. There is no \textit{a priori} reason
to expect the infrared fixed point to remain perturbative through this window,
although arguments to this effect have been advanced \cite{Gardi:1998ch}.

If the theory is formulated in the continuum and a running coupling
$\overline{g}^{2}(L)$ is defined at some length scale $L$, we have $L
(\partial / \partial L) \overline{g}^{2}(L) =
\overline{\beta}\left(\overline{g}^{2}(L)\right)$, where
\beq \label{beta}
\overline{\beta}\left(\overline{g}^{2}(L)\right) = b_{1} \overline{g}^{4}(L)
+ b_{2} \overline{g}^{6}(L) + b_{3} \overline{g}^{8}(L)+ b_{4}
\overline{g}^{10}(L) + \cdots\ .
\eeq
For the case of $\text{SU}(3)$, the first two (scheme-independent) coefficients are
\beq\label{univbeta}
\textstyle
b_1\!=\!\frac{2}{(4\pi)2} \left[  11 - \frac{2}{3}  N_f \right] ,
b_2\!=\!\frac{2}{(4\pi)4} \left[ 102 - \frac{38}{3} N_f \right] .
\eeq
The next two coefficients depend on the renormalization scheme. In the
$\overline{\text{MS}}$ scheme, they are given by \cite{Ritbergen:1997fk}
\beq \label{eq:b3}
b_3^{\overline{\text{MS}}} = \frac{2}{(4\pi)^6}\left[ \frac{2857}{2}
- \frac{5033}{18}N_f + \frac{325}{54}N_f^2 \right]\ ,
\eeq and
\beq
b_4^{\overline{\text{MS}}} = \frac{2}{(4\pi)^8} \left( 29243.0 - 6946.30 N_f + 405.089 N_f^2 + 1.49931 N_f^3\right) \ .
\eeq
For $N_f$ close to $33/2$, the two-loop infrared fixed point value $
\overline{g}^2_{*}$ is very small, and therefore corrected very little by the
higher order terms.

If the loop expansion is reliable to estimate $\overline{g}^2_{*}$, other
quantities can also be estimated. An example is the (scheme-dependent) parameter
$\gamma$ governing the approach to the fixed point. If the beta function is
linearized in the vicinity of the fixed point,
\beq 
\overline{\beta}\left(\overline{g}^{2}(L)\right) \simeq
\gamma \left[\overline{g}^2_{*}- \overline{g}^2(L) \right] \ , \eeq
then as $L \rightarrow \infty $, the approach to the fixed point from either
side is given by
\beq \label{eq:anomdim} \overline{g}^2(L)\rightarrow \overline{g}^2_{*} -
\frac{\text{const}}{L^{\gamma}} \ . \eeq

For $N_f = 12$, there is a two-loop infrared fixed point at $\overline{g}_*^2
\simeq 9.48 $, corrected to $\simeq 5.47$ at three loops in the
$\overline{\text{MS}}$ scheme, and to $\simeq 5.91$ at four loops. The
critical exponent is then $\gamma = 0.36$ at two loops, and in the $\overline{\text{MS}}$ scheme is 
given by $\gamma = 0.296$ at three loops and $\gamma = 0.282$ at four loops. The convergence of the loop expansion is not
guaranteed, but the fact that the expansion parameter at the fixed point
$\overline{g}_*^2/4\pi$ is relatively small suggests that it could be
reliable, and therefore that $N_f = 12$ lies inside the conformal window. For
$N_f = 8$, there is no two-loop infrared fixed point. A fixed point can
appear at three loops and beyond in some schemes, but its scheme dependence and typically large
value means that there is no reliable evidence for an infrared fixed point
accessible in perturbation theory.  A nonperturbative study is essential.

%%%%%%%%%%%%%%%%%%%%%%%%%%%%%%%%%%%%%%%%%%%%%%%%%%%%%%%%%%%%%%%%%%%%%%%%%%%%%%
\subsection{An upper bound on $N_f^\text{c}$}
%%%%%%%%%%%%%%%%%%%%%%%%%%%%%%%%%%%%%%%%%%%%%%%%%%%%%%%%%%%%%%%%%%%%%%%%%%%%%%

We next review a conjectured inequality which leads to an upper bound on
$N_f^\text{c}$ \cite{Appelquist:1999hr}. For any asymptotically free theory, the
thermodynamic free energy may be computed perturbatively as $T \rightarrow
\infty$, approaching the (free) Stefan-Boltzman expression
$-(\pi^{2}T^4/90)f_{\text{UV}}$, with $f_{\text{UV}} = [N_B + (7/8)4 N_F]$, where $N_B$ and
$N_F$ are the numbers of (underlying) bosonic fields and four-component Dirac
fields. Similarly, as $ T \rightarrow 0$, if the effective low energy theory
is infrared free, the free energy approaches the expression
$-(\pi^{2}T^4/90)f_{\text{IR}}$, where $f_{\text{IR}}$ counts the (massless) infrared degrees of
freedom in the same way. The conjectured inequality is simply $f_{\text{IR}} \leq
f_{\text{UV}}$.

For a nonsupersymmetric $\text{SU}(N)$ theory with $N_f$ massless Dirac fermions in
the fundamental representation, $f_{\text{UV}} = 2(N^2 - 1) + (7/8)4 N N_f$. If this
theory is in the chirally broken phase at zero temperature, then $f_{\text{IR}}$
simply counts the number of Goldstone bosons: $f_{\text{IR}} = N_f^2 -1$.  The
inequality then gives $N_f^\text{c} \leq 4\sqrt{N^2 - 16/81}$.  For $\text{SU}(3)$, this is
consistent with $N_f = 12$ being within the conformal window.

It is interesting to note that for a supersymmetric $\text{SU}(N)$ gauge
theory with $N_f$ massless Dirac fermions in the fundamental representation,
where $N_f^\text{c}$ denotes the transition point between the phase with infrared
conformal symmetry and the free-magnetic phase, the same inequality gives
$N_f^\text{c} \leq (3/2) N$, a limit precisely saturated by the result from duality
arguments \cite{Seiberg:1994pq}.  It is natural to ask to what extent the
inequality is also saturated in the nonsupersymmetric case.

%%%%%%%%%%%%%%%%%%%%%%%%%%%%%%%%%%%%%%%%%%%%%%%%%%%%%%%%%%%%%%%%%%%%%%%%%%%%%%
\subsection{Other studies}
%%%%%%%%%%%%%%%%%%%%%%%%%%%%%%%%%%%%%%%%%%%%%%%%%%%%%%%%%%%%%%%%%%%%%%%%%%%%%%

Finally, we note that several groups over the years have attempted to
determine the value of $N_f^\text{c}$, as well as the nature of the transition as
$N_f \rightarrow N_f^\text{c}$ from below and features of the bound-state spectrum in this
limit, by studying continuum Schwinger-Dyson (SD) equations
\cite{Appelquist:1997fp,Miransky:1996pd,Kurachi:2006ej}. Some truncation of
the SD equations must be adopted. It is then assumed that the infrared behavior is
governed by an infrared fixed point appearing in the perturbative beta function, 
and solutions corresponding to broken chiral symmetry are sought.
This leads to a value for $N_f^\text{c}$ slightly below $4N$, approaching it in
the large-$N$ limit, as well as information about the theory in the broken
phase near the transition. The reliability of these results is not clear,
however, since higher order effects are not obviously small.

%%%%%%%%%%%%%%%%%%%%%%%%%%%%%%%%%%%%%%%%%%%%%%%%%%%%%%%%%%%%%%%%%%%%%%%%%%%%%%
\section{The Schr\"odinger Functional Formalism}\label{sec:sf}
%%%%%%%%%%%%%%%%%%%%%%%%%%%%%%%%%%%%%%%%%%%%%%%%%%%%%%%%%%%%%%%%%%%%%%%%%%%%%%

%%%%%%%%%%%%%%%%%%%%%%%%%%%%%%%%%%%%%%%%%%%%%%%%%%%%%%%%%%%%%%%%%%%%%%%%%%%%%%
\subsection{Introduction}\label{sec:sfintro}
%%%%%%%%%%%%%%%%%%%%%%%%%%%%%%%%%%%%%%%%%%%%%%%%%%%%%%%%%%%%%%%%%%%%%%%%%%%%%%

The Schr\"odinger functional is the partition function describing the quantum
mechanical evolution of a system from a prescribed state at time $t = 0$ to
another state at time $t = T$ in a spatial box of size $L$ with periodic
boundary conditions \cite{Luscher:1992an,Sint:1993un,Bode:2001jv}.  Dirichlet
boundary conditions are imposed at $t = 0$ and $t = T$ where $T$ is $O(L)$.
They are chosen such that the minimum-action configuration is a constant
chromo-electric background field of strength $O(1/L)$. This can be implemented
in the continuum \cite{Luscher:1992an} or with lattice regularization
\cite{Bode:1999sm}.  In either case, by considering the response of the system
to small changes in the background field, a gauge invariant running coupling
can be defined, valid for any coupling strength.

The Schr\"odinger functional can be represented as the path integral
\beqs
\lefteqn{{\cal{Z}}[W,\zeta, \overline{\zeta}; W^\prime,\zeta^\prime,
\overline{\zeta}^\prime] =} \\*
&& \int [DA  D\psi D\overline{\psi}]
e^{-S_G(W,W^\prime) - S_F(W,W^\prime,\zeta,\overline{\zeta},\zeta^\prime,
\overline{\zeta}^\prime)}, \nonumber
\eeqs
where $A$ is the gauge field and $\psi$, $\overline{\psi}$ are the
fermion fields.  $W$ and $W^\prime$ are the (fixed) boundary values of the
gauge fields, and $\zeta, \overline{\zeta}, \zeta^\prime,
\overline{\zeta}^\prime$ are the boundary values of the fermion fields at
$t=0$ and $t=T$, respectively.

Although the Schr\"odinger functional can be formulated completely in the continuum, from
here on we will introduce a Euclidean spacetime lattice.  The quantity $S_G$ is
chosen to be the standard Wilson gauge action \cite{Wilson:1974sk} with a
boundary improvement counterterm:
 \beqs\label{gaugeact}
 \lefteqn{S_G = -\frac{\beta}{N} \sum_{P} w(P)  \textrm{Re}\ \textrm{Tr}\ U_P} \\*
 && = -\frac{\beta}{4N} a^4 \sum_{x} \textrm{Tr} \hat{F}_{\mu \nu} \hat{F}^{\mu \nu} - \frac{\beta}{4N} (1-c_t) a^5 \sum_{x} [\delta(t - 0) + \delta(t - T)] \mathrm{Tr}\ \hat{F}_{0\nu} \hat{F}^{0\nu} + \mathcal{O}(a^6), \nonumber
 \eeqs
 where Tr represents a color trace, $a$ is the lattice spacing, and $\beta
\equiv 2N / g_0^2$ with $g_0$ the lattice coupling constant. The improvement
coefficient $w(P) = c_t$ when multiplying timelike plaquettes which
intersect the Dirichlet boundaries, and is equal to $1$ elsewhere.  For this computation we set $c_t$ equal to its value as determined in one-loop lattice perturbation theory
\cite{Heller:1997pn},
\beq
 c_t = 1 + g_0^2 [-0.08900(5) + 0.00474(1) N_f].
 \eeq

The operator $\hat{F}_{\mu \nu}$ is defined similarly to the continuum field
strength tensor,
\beq
\hat{F}_{\mu \nu} \equiv \Delta^f_{\mu} A_\nu - \Delta^f_{\nu} A_\mu + [A_\mu, A_\nu]
\eeq
with $\Delta^f_{\mu} g(x) \equiv (g(x+a\hat{\mu}) - g(x))/a$ the discrete
forward derivative operator.  If we take the continuum limit of the action
(\ref{gaugeact}), we recover the standard Yang-Mills action.  The sum over
plaquettes $P$ may be expanded out in terms of individual gauge links:
\beq
 \sum_P \textrm{Re}\ \textrm{Tr}\ U_P = \sum_{x} \sum_{\mu < \nu} \textrm{Re}\ \textrm{Tr} \left[U_\nu(x) U_\mu(x+\hat{\nu}) U^\dagger_\nu(x+\hat{\mu}+\hat{\nu}) U^\dagger_\mu(x+\hat{\mu})\right].
 \eeq

For the fermionic action, we use the staggered approach as in Ref.
\cite{Heller:1997pn}, which reduces the 16 doubler species of a naively
discretized fermion field to 4 degrees of freedom.  In the continuum limit,
a single staggered fermion field can be interpreted as four degenerate Dirac fermion fields.  For $N_f$ divisible by four, the total fermionic action $S_F$ is
then given by
\beq
 S_F =   \sum_{N_f/4} S_f,
 \eeq
where $S_f$ is the fermion action for a single staggered field as in \cite{Heller:1997pn},
\beq
S_f = \frac{1}{2} \sum_{x, \mu} \eta^\mu(x) \overline{\chi}(x) \left[ U_\mu (x) \chi(x+\hat{\mu}) - U_\mu^\dagger (x-\hat{\mu}) \chi(x-\hat{\mu}) \right],
\eeq
with $\eta^\mu$ the usual staggered phase factor $\eta^\mu = (-1)^{\sum_{\nu < \mu} x_\nu}$.

Without affecting the action, the spatial-periodicity
condition can be generalized to include a phase rotation on the fermion fields
at the spatial boundaries,
\beq
\chi(x+L\hat{k}) = e^{i\theta_k} \chi(x),\ \ \overline{\chi}(x+L\hat{k}) = \overline{\chi}(x) e^{-i\theta_k},
\eeq
where $k$ runs over all of the spatial directions.  Imposing a nonzero value
on the $\theta_k$ has been shown in QCD to improve the ratio of the largest
and smallest eigenvalues of the Dirac matrix \cite{Sint:1996ys}, which in turn
improves computational speed.  However, this result is based on
nonperturbative tuning, and is not guaranteed to extend to the theories we
are considering.  We therefore set $\theta_k = 0$ for simplicity.

%%%%%%%%%%%%%%%%%%%%%%%%%%%%%%%%%%%%%%%%%%%%%%%%%%%%%%%%%%%%%%%%%%%%%%%%%%%%%%
\subsection{Temporal boundary values and definition of the Schr\"{o}dinger functional coupling}\label{sec:sfdef}
%%%%%%%%%%%%%%%%%%%%%%%%%%%%%%%%%%%%%%%%%%%%%%%%%%%%%%%%%%%%%%%%%%%%%%%%%%%%%%

The gauge boundary values $W, W^\prime$ are chosen such that the minimum-action configuration is a constant chromoelectric field whose magnitude is of
$O(1/L)$ and is controlled by a dimensionless parameter $\eta$.  The Schr\"odinger functional (SF)
running coupling is then defined in terms of the response of the action to
variations in $\eta$.  The setup is as follows: we take for the particular
boundary values of the gauge fields
\beq
\begin{cases}
U_k(\vec{x}, t=0) &= \exp (a C_k)\\
U_k(\vec{x}, t=T) &= \exp (a C_k')\end{cases}
\eeq
where the $C_k, C_k'$ are spatially constant and abelian,
\beq
C_k = \frac{i}{L} \left(\begin{array}{ccc}
\phi_1&&\\ &\phi_2 &\\ &&\phi_3\end{array}\right),\
C_k' = \frac{i}{L} \left(\begin{array}{ccc}
\phi_1'&&\\ &\phi_2'&\\ &&\phi_3'\end{array}\right).
\eeq

Classically, boundary conditions of this form lead to a constant
chromoelectric background field, with field strength proportional to $1/L$.
We adopt the particular set of boundary values
\beq
\left\{\begin{array}{lll}
\phi_1 = \eta - \frac{\pi}{3} &\hspace{5mm} & \phi_1' = -\eta - \pi \\
\phi_2 = -\frac{1}{2} \eta & \hspace{5mm} & \phi_2' = \frac{1}{2} \eta + \frac{\pi}{3} \\
\phi_3 = -\frac{1}{2} \eta + \frac{\pi}{3} & \hspace{5mm} & \phi_3' = \frac{1}{2} \eta + \frac{2\pi}{3}
\end{array}\right.,
\eeq
which are chosen to ensure that the background field is a stable solution to
the classical field equations for small $\eta$ \cite{Luscher:1993gh}.

With the boundary conditions fixed in this way, the SF running coupling
$\overline{g}^{2}(L,T)$ is defined by taking
\beq
\label{eq:dSdeta}
\frac{k}{\overline{g}^{2}(L,T)} = \left.
  - \frac{\partial}{\partial \eta} \log \cal{Z}
\right|_{\eta = 0}\ ,
\eeq
where
\beq
\label{eq:k}
k = 12 \left( \frac{L}{a} \right)^2 \left[
  \sin\left( \frac{2\pi a^2}{3LT}\right)
  + \sin\left( \frac{\pi a^2}{3LT} \right)
\right]\  .
\eeq
The factor $k$ is chosen so that $\overline{g}^{2}(L,T)$ equals the bare
coupling at tree level. In general, $\overline{g}^{2}(L,T)$ can be thought of
as the response of the system to small changes in the background
chromo-electric field.

The fermionic Dirichlet boundary values $\zeta, \overline{\zeta}, \zeta',
\overline{\zeta'}$ are subject only to multiplicative renormalization for staggered fermions
\cite{Sommer:2006qf}.  As we are free to choose these values, we take them
equal to zero, simplifying the calculation.

The staggered approach to discretization of fermions can be thought of as
splitting the 16 degrees of freedom of a single spinor over a $2^4$ hypercube
of lattice sites.  This framework makes it evident that staggered
simulations require an even number of lattice sites in each direction.  Thus with periodic boundary conditions in space, the spatial extent $L/a$ of the lattice must be even.  However, in the Schr\"odinger functional formalism, the Dirichlet boundaries in the time direction require an odd temporal extent $T/a$ in order for the number of lattice sites to be even, since the sites located at $t=0$ and $t=T$ are distinct.

As a result, one cannot simulate with $T=L$, only with $T=L \pm a$.  In the
continuum limit $T=L$ is recovered, but at a finite lattice spacing this
results in the introduction of $O(a)$ lattice artifacts into observables.
This is particularly undesirable, since staggered fermion simulations
contain bulk artifacts at $O(a^2)$ and higher.  Fortunately,
simulating at $T = L \pm a$ and averaging over the observed values eliminates
this effect \cite{Heller:1997pn}.  We adopt this technique here, defining the
central observable
\beq
\label{eq:bulk_averaging}
\frac{1}{\overline{g}^{2}(L)} = \frac{1}{2} \left[
  \frac{1}{\overline{g}^{2}(L,L-a)} + \frac{1}{\overline{g}^{2}(L,L+a)}
\right],
\eeq
which depends on only one large distance scale $L$. To be more explicit, this
running coupling can be written as $\overline{g}^2(\beta, L/a)$ where $\beta
\equiv 2N / g_0^2$.  From this point on we will fix $N = 3$, and so $\beta = 6 / g_0^2$.

%%%%%%%%%%%%%%%%%%%%%%%%%%%%%%%%%%%%%%%%%%%%%%%%%%%%%%%%%%%%%%%%%%%%%%%%%%%%%%
\subsection{Schr\"odinger functional perturbation theory}\label{sec:sfpt}
%%%%%%%%%%%%%%%%%%%%%%%%%%%%%%%%%%%%%%%%%%%%%%%%%%%%%%%%%%%%%%%%%%%%%%%%%%%%%%

The SF coupling $\overline{g}^2(L)$ has been
normalized to give the bare lattice coupling $g_0^2$ at tree level. With the
lattice as a regulator, it can be expanded as a power series in $g_0^2$ with
coefficients depending on $a/L$. By rearranging this series in terms of a
coupling defined at an arbitrary scale and setting to zero terms that vanish
as $a \rightarrow 0$, a continuum beta function can be defined. Its
perturbation expansion leads to the universal coefficients $b_1$ and $b_2$ of
Eq.~(\ref{univbeta}) at the one- and two-loop levels.

The three-loop, scheme-dependent coefficient has been computed in the Schr\"odinger functional  scheme by combining the two-loop perturbative computation of the SF running
coupling in lattice perturbation theory with a similar lattice computation of
the $\overline{\text{MS}}$ coupling constant \cite{Bode:1999sm}. The result is
\beq
b_3^\text{SF} = b_3^{\overline{\text{MS}}} + \frac {b_{2}c_2}{4\pi}
- \frac{b_{1}(c_3 - c_2^{2})} {16 \pi^2}\ ,
\eeq
where $b_3^{\overline{\text{MS}}}$ is given by Eq.~(\ref{eq:b3}) with $c_2 = 1.256 + 0.040
N_f$ and $c_3 = c_2^2 + 1.197(10) + 0.140(6) N_f
- 0.0330(2)N_f^2$. The perturbative behavior discussed in Sec.~\ref{sec:pt}, based on
the $\overline{\text{MS}}$ scheme, is qualitatively unchanged by the
modification of the three-loop coefficient. For $N_f = 12$, the three-loop SF
coupling has a fixed point at $\overline{g}_\star^2 \approx 5.18$, compared
with $\overline{g}_\star^2 \approx 5.47$ at three loops in the
$\overline{\text{MS}}$ scheme. At three loops in the SF scheme, the value of the critical exponent Eq.~(\ref{eq:anomdim}) is $\gamma = 0.286$.

The four-loop coefficient in the SF scheme has not yet been computed. But the
fact that in the $\overline{\text{MS}}$ scheme the four-loop correction shifts
the fixed point by less than $10\%$ from its three-loop value suggests that
the same may be true in the SF scheme. This and the relative smallness of the
expected loop expansion parameter $O(\overline{g}^2(L)/(4\pi^{2}))$ at the
fixed point indicates that perturbation theory could be reliable to describe
infrared behavior for $N_f = 12$, and that the infrared fixed point might
truly exist. For $N_f = 8$, since the universal one- and two-loop coefficients are both
positive, there is no reliable, perturbative evidence for the existence of an
infrared fixed point. As observed in Sec.~\ref{sec:pt}, a nonperturbative study is
essential.

In perturbation theory, the SF running coupling behaves just like the running coupling defined in other, more familiar ways. Its behavior is identical through two loops, and then qualitatively similar at three loops and beyond. Other definitions of the running coupling are based on Green functions of local operators or quantities such as the Wilson loop, while the Schr\"odinger functional and the associated running coupling dependence on a background field act across the entire lattice. This definition of the running coupling is nonperturbative, but its relation to other nonperturbative definitions in the strong-coupling regime is not yet clear. Nevertheless, the SF running coupling is adequate for our purposes: to distinguish between conformal and confining behavior in the infrared.

%%%%%%%%%%%%%%%%%%%%%%%%%%%%%%%%%%%%%%%%%%%%%%%%%%%%%%%%%%%%%%%%%%%%%%%%%%%%%%
\section{Lattice Simulations}\label{sec:lat}
%%%%%%%%%%%%%%%%%%%%%%%%%%%%%%%%%%%%%%%%%%%%%%%%%%%%%%%%%%%%%%%%%%%%%%%%%%%%%%

%%%%%%%%%%%%%%%%%%%%%%%%%%%%%%%%%%%%%%%%%%%%%%%%%%%%%%%%%%%%%%%%%%%%%%%%%%%%%%
\subsection{Setup and Procedure}\label{sec:latsetup}
%%%%%%%%%%%%%%%%%%%%%%%%%%%%%%%%%%%%%%%%%%%%%%%%%%%%%%%%%%%%%%%%%%%%%%%%%%%%%%

To measure the running coupling on the lattice, we generate an ensemble of
gauge configurations distributed with the appropriate weighting by the
Euclidean action.  The running coupling is then computed as an
average over this ensemble.  Simulations are performed using the MILC code
\cite{MILC}, with some customization.  Evolution of the gauge configurations
is accomplished using the hybrid molecular dynamics (HMD) approach, with
the fermionic contribution included using the R algorithm \cite{Gottlieb:1987mq}.
Trajectories are taken to be of unit length, while the step size $\Delta \tau$
of the MD integrator is varied.  The R algorithm is known to introduce errors
of $O((\Delta \tau)^2)$ into observables; we discuss this effect along with
other sources of error below.

Sets of gauge configurations are generated at a fixed box size $L/a$ and fixed
bare coupling $\beta$.  For each ($\beta, L/a$), two
independent ensembles are created at $T/a = L/a \pm 1$, and then averaged
together as in Eq.~(\ref{eq:bulk_averaging}). The data are collected over a wide
range of $\beta$ values and for $6 \leq L/a \leq 20$, in order to capture the
evolution of $\overline{g}^2(L)$ over a large range of scales.  It is important to note that in the range of $\beta$ values employed, for both $N_f = 8$ and $N_f = 12$, there is no evidence for a bulk phase transition.  We have explored this issue by examining the plaquette time series within this range and at lower values of $\beta$.  At lower values, we have indeed found evidence for a bulk phase transition.  These lower values are, however, well separated from the minimum $\beta$ shown in our data tables and used in our analysis.  There is no such evidence in the range of $\beta$ values employed here.

Simulations were also performed at $L/a = 4$, but these values were not used in
our analysis.  Examination reveals large lattice-artifact corrections
in the $L/a = 4$ data.  Their presence is not unexpected,
particularly on the smaller $4^3\times 3$ lattices; with only a single lattice site between
the Dirichlet boundaries, the $O(a)$ boundary operators appearing in Eq.~(\ref{gaugeact}) overlap.  In addition, the $4^3 \times 3$ lattices fail to satisfy the
conditions of the ``stability theorem" of L\"{u}scher \textit{et al.}, meaning that
the background chromoelectric field being expanded around may not be an absolute
minimum of the action \cite{Luscher:1992an}.  This also precludes the use of data from $4^3 \times 5$ lattices, since although they satisfy the stability criterion, without the $4^3 \times 3$ lattices we cannot use the averaging procedure described in Sec.~\ref{sec:sfdef} \footnote{There has been some ongoing work by Perez-Rubio \textit{et al.}\ attempting to remove the bulk $O(a)$ artifact directly through a counterterm, rather than via the averaging procedure, which might allow useful measurement of the SF running coupling on the $4^3 \times 5$ lattices \cite{PerezRubio:2007qz}. }.

Since updating of the gauge fields is accomplished locally, while the running
coupling is simulated on the scale of the box size $L$, a large
number of updates is required to generate statistically independent
values of $\overline{g}^2(L)$.  To remove statistical bias from our
results, we collect a large number of gauge configurations at each point in parameter space, ranging from
$20 000$ to $160 000$ MD trajectories with a greater number required for
measurements at stronger coupling.  These run lengths are established based on our
analysis of autocorrelations, discussed in Appendix~\ref{app:error}.

%%%%%%%%%%%%%%%%%%%%%%%%%%%%%%%%%%%%%%%%%%%%%%%%%%%%%%%%%%%%%%%%%%%%%%%%%%%%%%
\subsection{Step scaling}\label{sec:latss}
%%%%%%%%%%%%%%%%%%%%%%%%%%%%%%%%%%%%%%%%%%%%%%%%%%%%%%%%%%%%%%%%%%%%%%%%%%%%%%

We are interested in mapping out the behavior of the running coupling over a
large range of scales, from the ultraviolet to the infrared.  Often, a
lattice simulation (i.e. a set of gauge configurations, generated using a
fixed set of parameters) is focused on measuring quantities at distance scales
lying between the lattice spacing $a$ and the box size $L$.  Our use of the
Schr\"odinger functional instead places the observable
$\overline{g}^2(L)$ {\it at the scale $L$}, eliminating the latter
restriction. However, the range of scales $L$ over which we can measure the
coupling strength with fixed lattice spacing $a$ before the computational
expense becomes prohibitive is still rather limited.  To achieve our
goal, therefore, we must measure the coupling using a wide range of
$a$ values, and then match these measurements together.  To accomplish this,
we use a procedure known as step scaling
\cite{Luscher:1991wu,Caracciolo:1994ed}.

Step scaling provides a systematic way to combine multiple lattice
measurements of the running coupling $\overline{g}^2(L)$ into a single
measurement of the continuum evolution of the coupling as the scale changes
from $L \rightarrow sL$, where $s$ is a scaling factor called the step size. In
a continuum setting, the evolution is described by the ``step-scaling function,"
\beq
\sigma(s, \overline{g}^2(L)) \equiv \overline{g}^2(sL),
\eeq
which can be thought of as a discrete version of the usual continuum evolution
described by the beta function.  In a lattice calculation, the
extracted step-scaling function will be a function also of $a/L$, which we must
extrapolate to the continuum:
\beq
\sigma(s,\overline{g}^2(L)) = \lim_{a\rightarrow 0}
\Sigma(s,\overline{g}^2(L),a/L).
\eeq

Step scaling is generically implemented on the lattice as follows.  First, an
initial value $u = \overline{g}^2(L)$ is chosen.  Several ensembles with
different values of $a/L$ are then generated, with $\beta$
tuned so that the coupling measured on each is equal to the chosen value,
$\overline{g}^2(L) = u$.  A second ensemble is generated at each $\beta$,
but with $L \rightarrow sL$.  The value of the coupling measured on this
larger lattice is exactly $\Sigma(s, u, a/L)$.  An extrapolation $a/L
\rightarrow 0$ can then recover the continuum value $\sigma(s, u)$.  Taking
$\sigma(s,u)$ to be the new starting value, we can then iterate this procedure
until we have sampled $\overline{g}^2(L)$ over a large range of $L$ values.
In practice we take $s=2$.

There is a natural caveat on the step-scaling procedure. In the limit
$a/L \rightarrow 0$ with $\overline{g}^2(L)$ fixed, $g_0^{2}(a/L)$ depends on
the short-distance behavior of the theory, and it is important that it remains
bounded so that it does not trigger a bulk phase transition. If asymptotic
freedom governs the short distance behavior, this is automatic since
$g_0^{2}(a/L) \rightarrow 1/\log(L/a)$. While this is our principal focus, the
existence of an infrared fixed point for the $N_f = 12$ theory will lead us to
consider also values of  $\overline{g}^2(L)$ lying above the fixed point. Then
$g_0^{2}(a/L)$ increases as $a \rightarrow 0$, with no evidence from our
simulations that it remains bounded and therefore that the
continuum limit exists. Nevertheless, one can consider small values of $a/L$
 providing that $g_0^{2}(a/L)$ remains
small enough not to trigger a bulk phase transition. We return to this point
in our discussion of the $N_f = 12$  simulation data.

%%%%%%%%%%%%%%%%%%%%%%%%%%%%%%%%%%%%%%%%%%%%%%%%%%%%%%%%%%%%%%%%%%%%%%%%%%%%%%
\subsection{Interpolating Functions}\label{sec:latint}
%%%%%%%%%%%%%%%%%%%%%%%%%%%%%%%%%%%%%%%%%%%%%%%%%%%%%%%%%%%%%%%%%%%%%%%%%%%%%%

Carrying out the above procedure directly can be expensive in computational
power since each tuning of $\beta$ may require several attempts. The procedure also severely limits the rate at which computations may be performed, since each simulation must be finished and the value of
$\sigma(s,u)$ extracted before the next iteration. We instead measure
$\overline{g}^2(L)$ for a limited set of values for $\beta$ and $L/a$, and then generate an interpolating function. This function is then used
to tune $\beta$ as described above, and renders the cost of extracting a
step-scaling function independent of the number of steps taken.

For any value of $L/a$ in our range, $\overline{g}^2(\beta, L)$ is a monotonically decreasing function of $\beta = 6/g_0^2$.  One procedure is simply to interpolate linearly between the available $\beta$ values for each $L/a$.  This works reasonably well in the perturbative region, reproducing the correct continuum perturbative running once the step-scaling procedure is carried out.  For stronger coupling, however, linear interpolation leads to some anomalies due to statistical fluctuations.  A better procedure is to use a smooth interpolating function fit to the data.

For our results reported in Ref. \cite{Appelquist:2007hu}, we fit $\overline{g}^2(\beta, L)$ to a single function based on a truncated Laurent series with poles at small, unphysical values of $\beta$, well below the simulation range.  Here we employ a set of interpolating functions, one for each $L/a$, focused directly on the lattice observable $1/\overline{g}^2(\beta, L/a)$.  Motivated by the fact that in lattice perturbation theory this quantity takes the form
%%%%%%%%%%%%%%%%%%%%%%%%%%%
\beq
\frac{1}{\overline{g}^2(\beta, L/a)} = \frac{1}{g_0^2} \left[ 1 + O(g_0^2) \right] = \frac{\beta}{6} \left[ 1 + O(\frac{1}{\beta})\right],
\eeq
%%%%%%%%%%%%%%%%%%%%%%%%%%%
we use a fit to $\overline{g}^2(\beta, L/a)$ at each $L/a$ as a function of $\beta$, with $n$-th order polynomial dependence on $g_0^2 = 6/\beta$:
%%%%%%%%%%%%%%%%%%%%%%%%%%%
\beq
\label{eq:interpolating_function}
\frac{1}{\overline{g}^2(\beta, L/a)} = \frac{\beta}{6} \left[ 1 - \sum_{i=1}^{n} c_{i,L/a} \left(\frac{6}{\beta}\right)^{i} \right].
\eeq
%%%%%%%%%%%%%%%%%%%%%%%%%%%
The order $n$ of the polynomial is varied with $L/a$ in order to achieve the optimal $\chi^2$ per degree of freedom when fitting to our data.  The values of the parameters with associated errors, determined by fits to the simulation data for both $N_f = 8$ and $N_f = 12$, are recorded in Appendix~\ref{app:data}.

This function is used for interpolation within the measured range, as a basis for the step-scaling procedure. It is based on empirical observation of the $g_0^2$ dependence of our observable, and is not meant to imply that perturbation theory is applicable to our nonperturbative, strong-coupling results.  More elaborate interpolating functions could be used, in particular, modeling explicitly the $L/a$ dependence or including nonanalytic terms in $g_0^2$, but such functional forms do not significantly alter the fit quality or the results of step scaling based on the collected data set.

%%%%%%%%%%%%%%%%%%%%%%%%%%%%%%%%%%%%%%%%%%%%%%%%%%%%%%%%%%%%%%%%%%%%%%%%%%%%%%
\subsection{Statistical and systematic errors}\label{sec:laterror}
%%%%%%%%%%%%%%%%%%%%%%%%%%%%%%%%%%%%%%%%%%%%%%%%%%%%%%%%%%%%%%%%%%%%%%%%%%%%%%

We account for numerous sources of statistical and systematic error in our analysis.  A detailed discussion of the estimation and/or elimination of these errors is given in Appendix~\ref{app:error}.  We conclude that potential systematic errors in our procedure are small compared to current statistical errors. Our final results for continuum running are therefore shown with only a statistical-error band.

We note that this conclusion differs from that of Ref.~\cite{Appelquist:2007hu}.  In that reference, statistical errors were estimated in a less sophisticated way, in particular, ignoring the accumulation of error over repeated step-scaling steps.  This led to very small statistical error bars.  In contrast, the systematic error as determined by uncertainty in the correct form of the continuum extrapolation was large, due primarily to the inclusion of values of $\overline{g}^2(L)$ computed on $L/a = 4$ volumes, which we now discard for reasons discussed in Sec.~\ref{sec:latsetup}.  Thus, with the more extensive analysis described here, statistical errors dominate rather than systematic.

%%%%%%%%%%%%%%%%%%%%%%%%%%%%%%%%%%%%%%%%%%%%%%%%%%%%%%%%%%%%%%%%%%%%%%%%%%%%%%
\section{Results}\label{sec:res}
%%%%%%%%%%%%%%%%%%%%%%%%%%%%%%%%%%%%%%%%%%%%%%%%%%%%%%%%%%%%%%%%%%%%%%%%%%%%%%

%%%%%%%%%%%%%%%%%%%%%%%%%%%%%%%%%%%%%%%%%%%%%%%%%%%%%%%%%%%%%%%%%%%%%%%%%%%%%%
\subsection {$N_f = 8$}
%%%%%%%%%%%%%%%%%%%%%%%%%%%%%%%%%%%%%%%%%%%%%%%%%%%%%%%%%%%%%%%%%%%%%%%%%%%%%%

The simulation data for $\overline{g}^2(L)$ as a function of $\beta$ and $L/a$ are displayed in Table~\ref{table:Nf8}. Each data point
is the average given by Eq.~(\ref{eq:bulk_averaging}), with the statistical
error in parentheses. The table ranges from $\beta = 4.55 \text{--} 192$ and
$L/a = 6 \text{--} 16$. The lower limit is chosen to insure that the lattice coupling is weak enough so as not to induce a bulk phase transition.  The upper limit is taken to be large so that we can check the agreement of our simulations with perturbation theory when the coupling is very weak. The final results reported here depend sensitively only on simulations below $\beta = 10$.  The data becomes more sparse with increasing $L/a$, reflecting the growing computational time involved. In particular, only a very limited
amount of $L/a = 20$ data, at very weak coupling, is available at $N_f = 8$, so we exclude these points from our analysis.  The $L/a = 10$ data is thus also excluded, since it cannot be used in step scaling at $s=2$ without the $L/a = 20$ points.  The listed
values of $\overline{g}^2(L)$ are perturbative ($\overline{g}^2(L)/4\pi \ll
1$) throughout much of the table, except for small $\beta$.

\begin{figure}
\includegraphics[width=150mm]{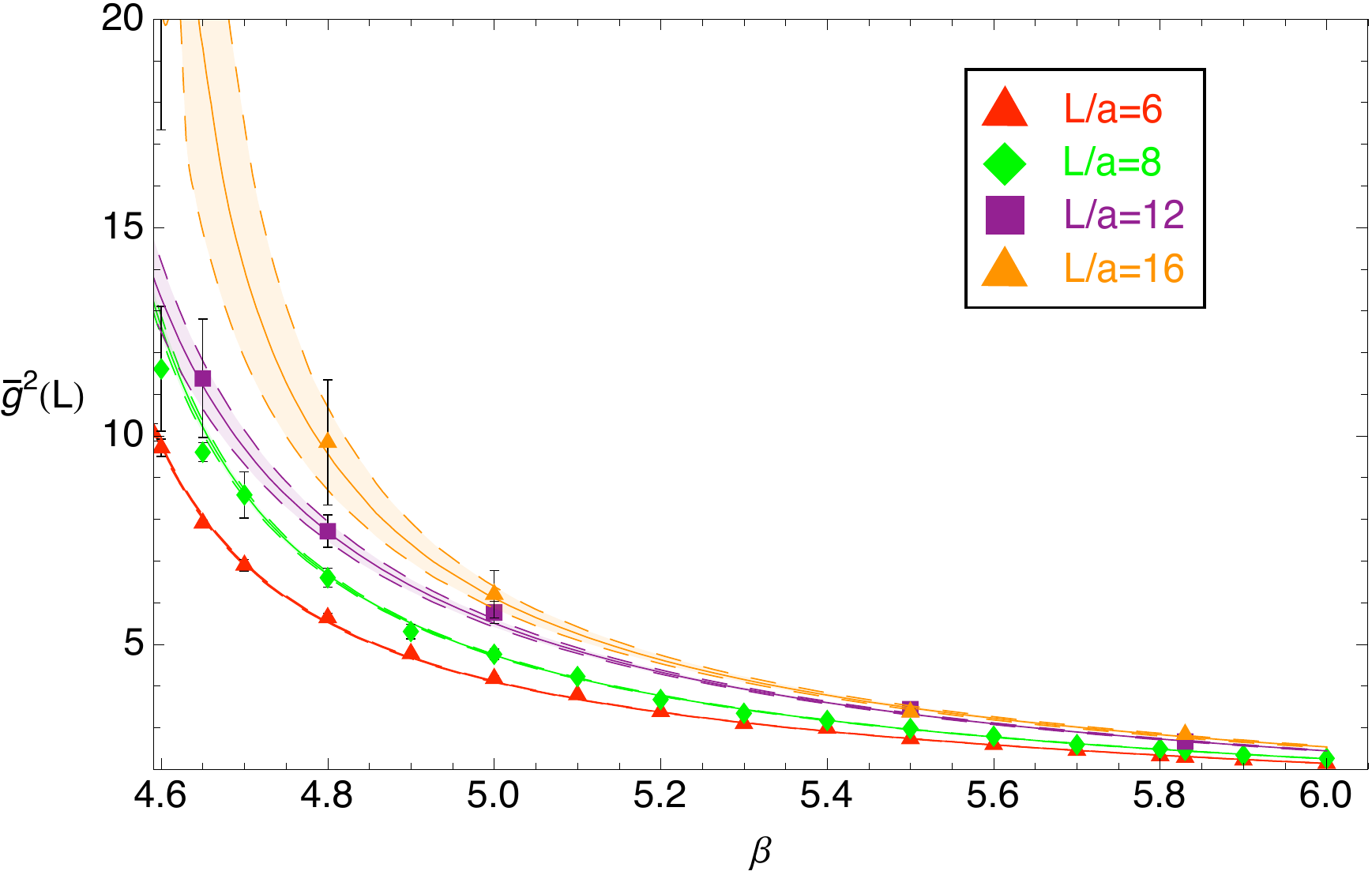}
\caption{\label{fig:dat-Nf8}Measured values $\overline{g}^2(L)$ versus $\beta$ for $N_f = 8$.  The interpolating curves shown represent the best fit to the data, using the functional form Eq.~(\ref{eq:interpolating_function}). The errors are statistical, derived as discussed in Appendix~\ref{app:error}.}
\end{figure}

In order to carry out the step-scaling procedure, we employ
the interpolating function of Eq.~(\ref{eq:interpolating_function}).
The resulting best-fit mean values and errors for the parameters at each $L/a$ are shown in Table~\ref{table:Nf8pms}. More details, including full covariance matrices, will be made available in the AIP's Electronic Physics Auxiliary Publication Service \cite{fitparameters}. In Fig.~\ref{fig:dat-Nf8},
data points are shown together with the interpolating functions for
$\overline{g}^2(L)$ as a function of $\beta$, for each of $L/a = 6, 8, 12, 16$.

\begin{figure}
\includegraphics[width=130mm]{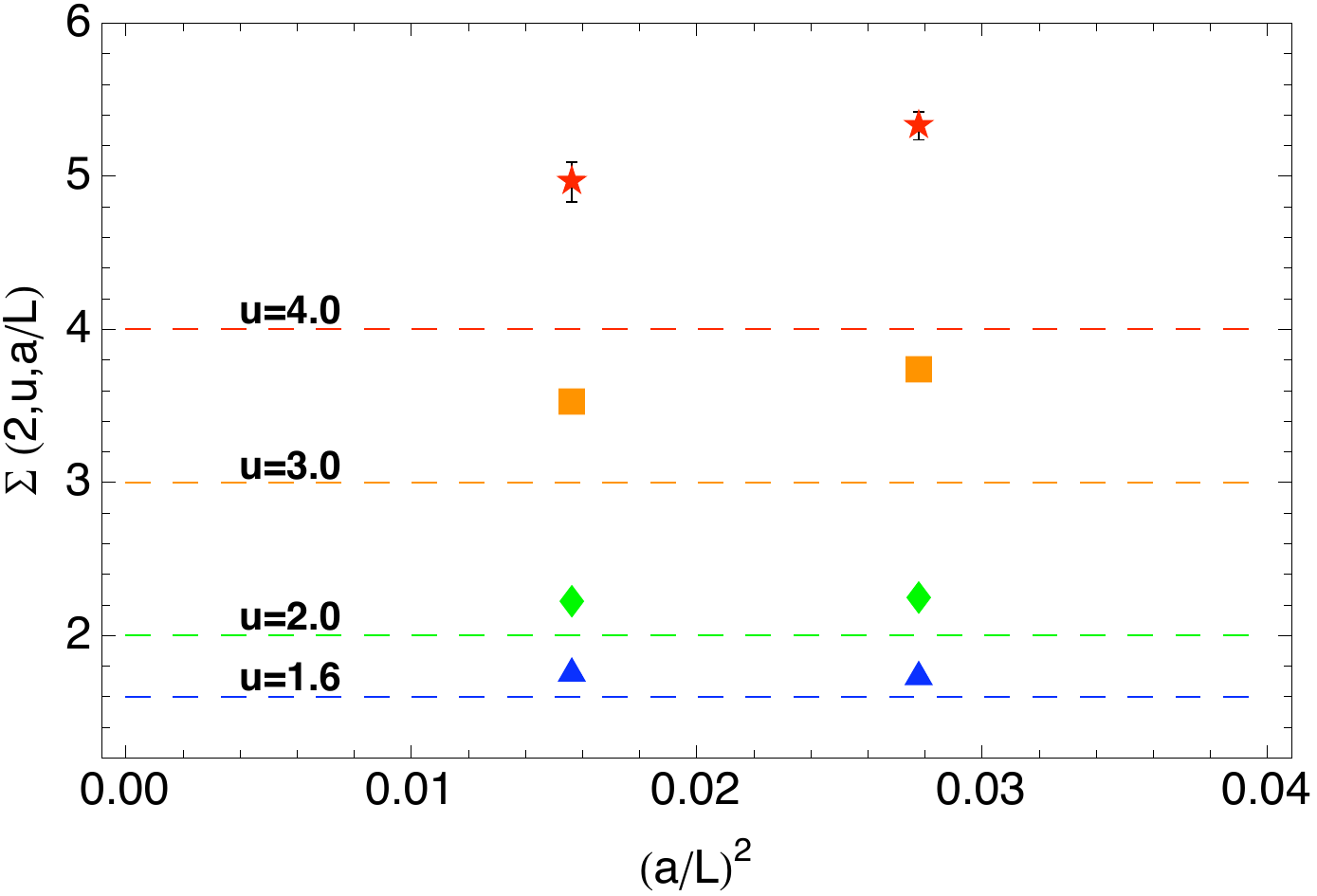}
\caption{\label{fig:extrap8} Step-scaling function $\Sigma(2,u,a/L)$ at various $u$, for each of the two steps $L/a = 6\rightarrow 12$ and $8 \rightarrow 16$ used in the $N_f = 8$ analysis.  Note that $\Sigma(2,u,a/L) > u$ in each case, with the difference increasing as $u$ increases.}
\end{figure}

We implement the step-scaling procedure and extrapolation to the continuum
as described in Sec.~\ref{sec:lat}. Figure~\ref{fig:extrap8} shows a typical continuum extrapolation from our 8-flavor data.  The points shown represent steps from $L/a = 6 \rightarrow 12$ and $8 \rightarrow 16$.  Constant extrapolation (a weighted average of the two points) is used since the lattice-artifact contributions to $\Sigma(2,u,a/L)$ are small compared to the statistical errors.  We have estimated the systematic error in this procedure and found that it is small compared to the statistical error; details of this analysis are provided
in Appendix~\ref{app:error}.

\begin{figure}
\includegraphics[width=150mm]{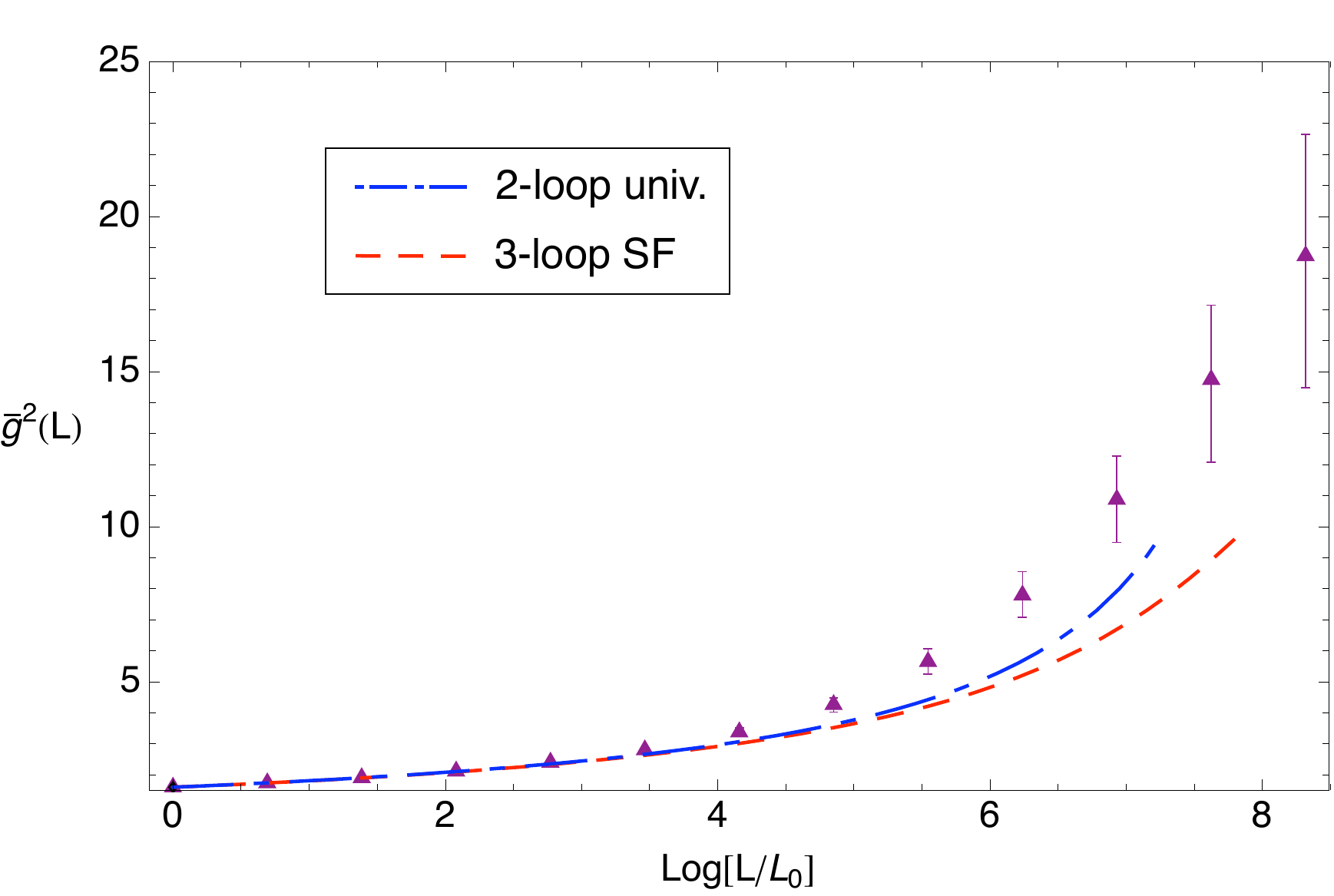}
\caption{\label{fig:ssf-Nf8}Continuum running for $N_f = 8$.  Purple points are derived by step-scaling using the constant continuum-extrapolation of Fig.~\ref{fig:extrap8}. The error bars shown are purely statistical, and are derived as described in Appendix~\ref{app:error}.  Two-loop and three-loop perturbation theory curves are shown for comparison.}
\end{figure}

Our results for the continuum running of $\overline{g}^2(L)$ are shown in
Fig.~\ref{fig:ssf-Nf8}. We take $L_0$ to be the scale at which $\overline{g}^2(L_0) =
1.6$, a relatively weak value. The points are shown for values of
$L/L_0$ increasing by factors of $2$. The (statistical) errors are derived as described in Appendix~\ref{app:error}. For comparison, the perturbative running of $\overline{g}^2(L)$
at two loops and three loops is shown up through $\overline{g}^2(L) \approx 10$ where perturbation theory is no longer expected to be accurate.
The results show that the coupling evolves according to perturbation theory up through $\overline{g}^2(L) \approx 4$, and then increases more rapidly, reaching values that clearly exceed typical estimates of
the strength required to trigger spontaneous chiral symmetry breaking \cite{Appelquist:1998rb}. The dynamical fermion mass is of order of the corresponding $1/L$, and since the coupling is strong here, the theory will
confine at roughly this distance scale. There is no evidence for an infrared fixed point or even an inflection point in the behavior of $\overline{g}^2(L)$.

%%%%%%%%%%%%%%%%%%%%%%%%%%%%%%%%%%%%%%%%%%%%%%%%%%%%%%%%%%%%%%%%%%%%%%%%%%%%%%
\subsection{$N_f = 12$}
%%%%%%%%%%%%%%%%%%%%%%%%%%%%%%%%%%%%%%%%%%%%%%%%%%%%%%%%%%%%%%%%%%%%%%%%%%%%%%

\begin{figure}
\includegraphics[width=150mm]{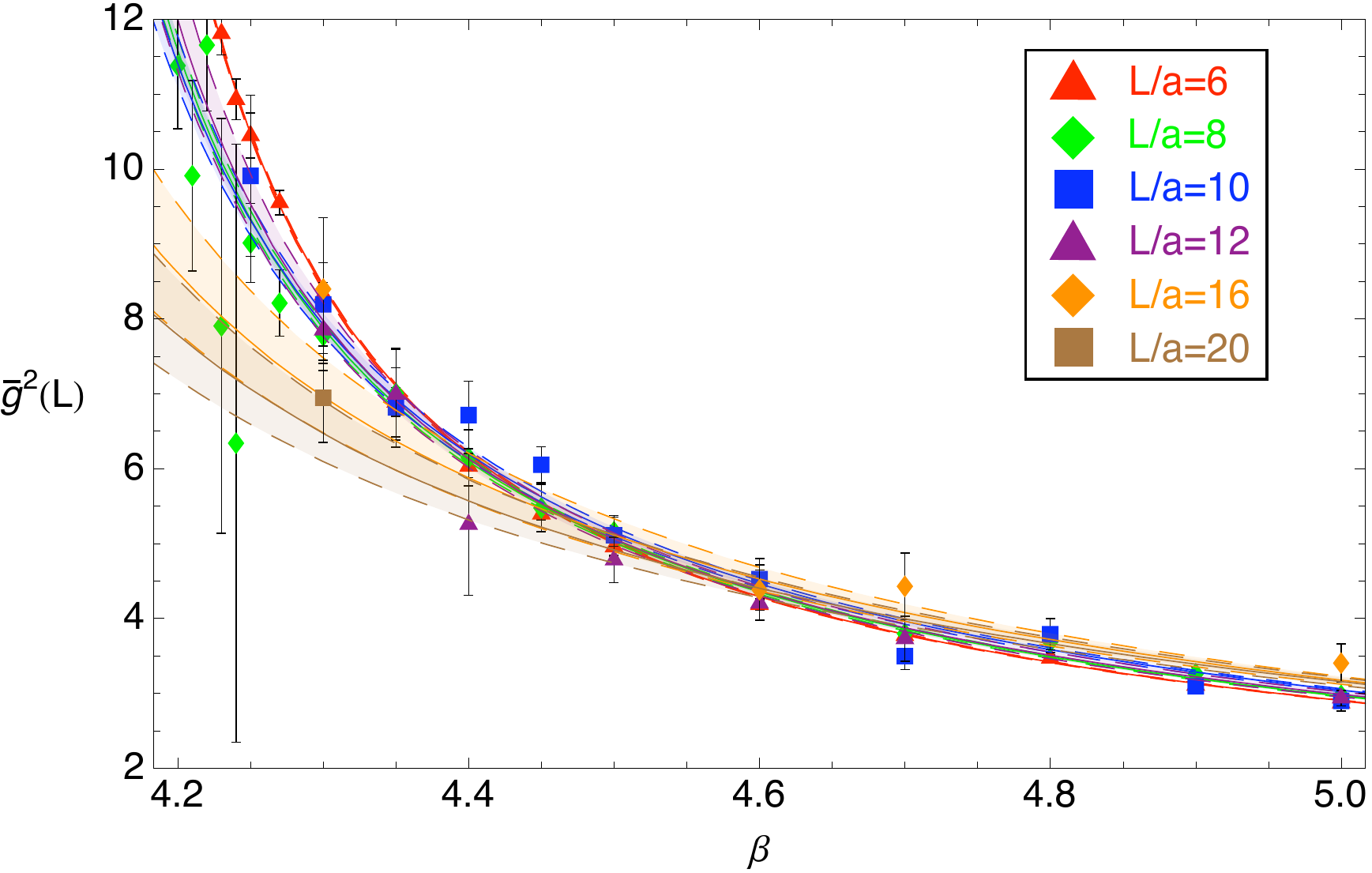}
\caption{\label{fig:dat-Nf12}Measured values $\overline{g}^2(L)$ versus $\beta$, $N_f = 12$.  The interpolating curves shown represent the best fit to the data, using the functional form of Eq.~(\ref{eq:interpolating_function}).}
\end{figure}

The simulation data for $\overline{g}^2(L)$ as a function of $\beta$ and $L/a$ are displayed in Table~\ref{table:Nf12}. As with $N_f =
8$, each data point is the average given by Eq.~(\ref{eq:bulk_averaging}),
with the estimated error in parentheses. The table ranges from $\beta = 4.2 \text{--} 192$ and $L/a = 6 \text{--} 20$. The lower limit on $\beta$ insures that the lattice coupling is weak enough so as not to induce a bulk phase transition.  As in the $N_f = 8$ case, the upper limit is taken to be large in order to explore agreement with perturbation theory, but data above $\beta = 10$ do not have significant influence on our analysis.  $L/a = 20$ data are included here and not in the $N_f = 8$ case because of concerns about the magnitude of the lattice artifact corrections, compared to the continuum running.  In the end, artifact corrections were found to be small compared to our statistical error.  Data become more sparse with increasing $L/a$, reflecting the growing computational cost involved.  The interpolating functional form Eq.~(\ref {eq:interpolating_function}) is
again employed, and the resulting best-fit mean values and errors of the parameters at each $L/a$ are shown in Table~\ref{table:Nf12pms}.
The full covariance matrix will be made available in the AIP's Electronic Physics Auxiliary Publication Service
\cite{fitparameters}. In Fig.~\ref{fig:dat-Nf12}, data
points are shown for
$\overline{g}^2(L)$ as a function of $\beta$, together with the interpolating functions for each of $L/a = 6, 8, 10, 12, 16,
20$.

The data and the interpolating curves already suggest the existence of an infrared
fixed point for $N_f = 12$.  For small $\beta$, the general trend is that $\overline{g}^2(L)$ decreases with increasing $L$. This behavior and the fact that for larger $\beta$, $\overline{g}^2(L)$ increases with increasing $L$, is reflected in a crossover behavior in the interpolating curves. 
We first implement the step-scaling procedure choosing an initial $u = \overline{g}^2(L)$
well below a possible fixed-point value so that a continuum limit is guaranteed to exist, as discussed in
Sec.~\ref{sec:latss}.

\begin{figure}
\includegraphics[width=130mm]{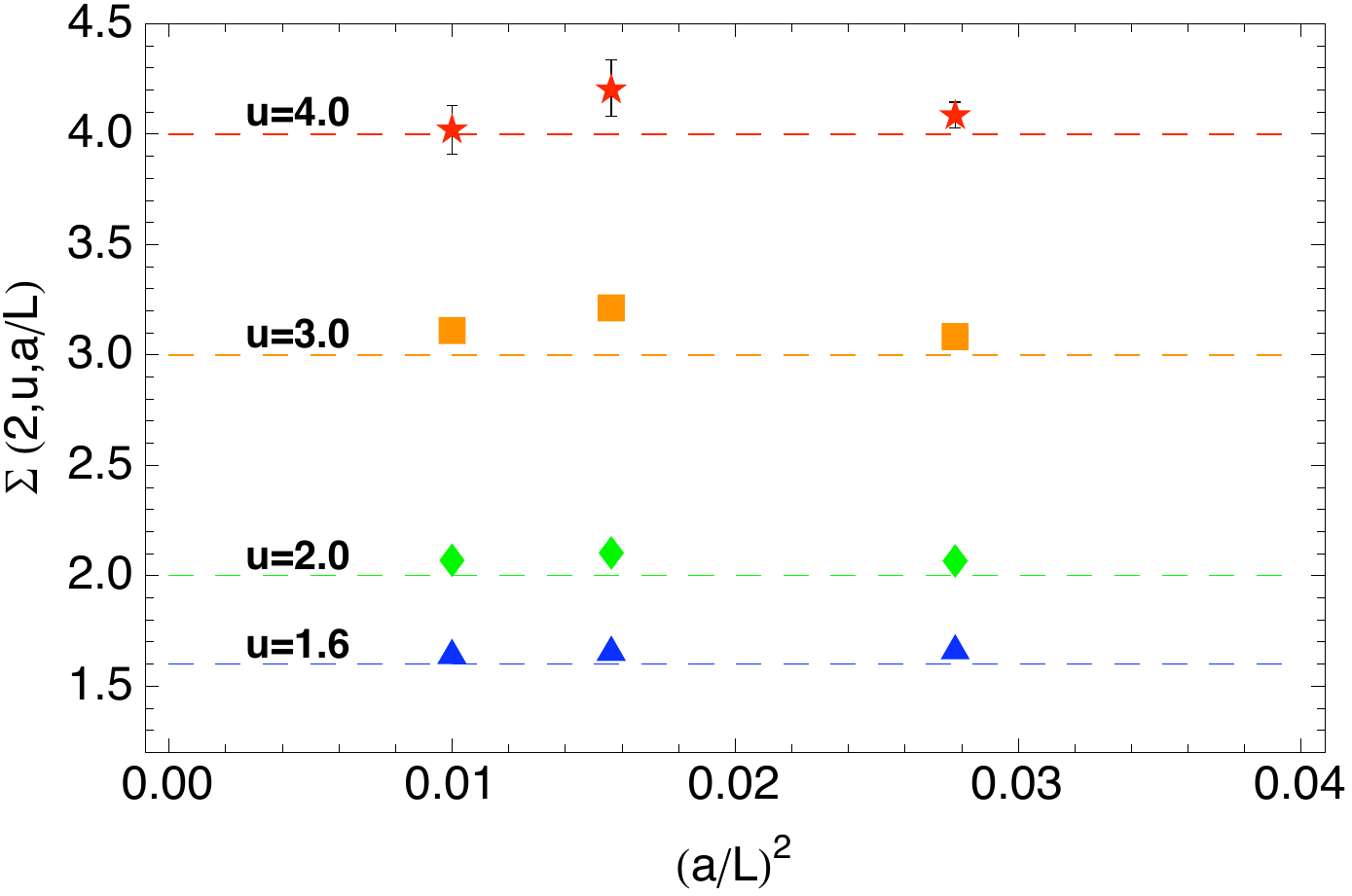}
\caption{\label{fig:extrap12}Step-scaling function $\Sigma(2,u,a/L)$ at various $u$, for each of the three steps $L/a = 6\rightarrow 12$, $8 \rightarrow 16$, $10 \rightarrow 20$ used in the $N_f = 12$ analysis. Note that $\Sigma(2,u,a/L) \rightarrow u$ as the starting coupling $u$ approaches the fixed point value.}
\end{figure}

A constant continuum extrapolation (a weighted average of the available values of $\Sigma(2,u,a/L)$) is again employed for each $u$. Now, since we have data at $L=20$, the extrapolation is a weighted average of three points corresponding to the steps $6 \rightarrow 12$, $8 \rightarrow 16$, and $10 \rightarrow 20$. Examples of such a continuum extrapolation are shown in Fig.~\ref{fig:extrap12}. The systematic error is again estimated to be small compared to the statistical error.

\begin{figure}
\includegraphics[width=150mm]{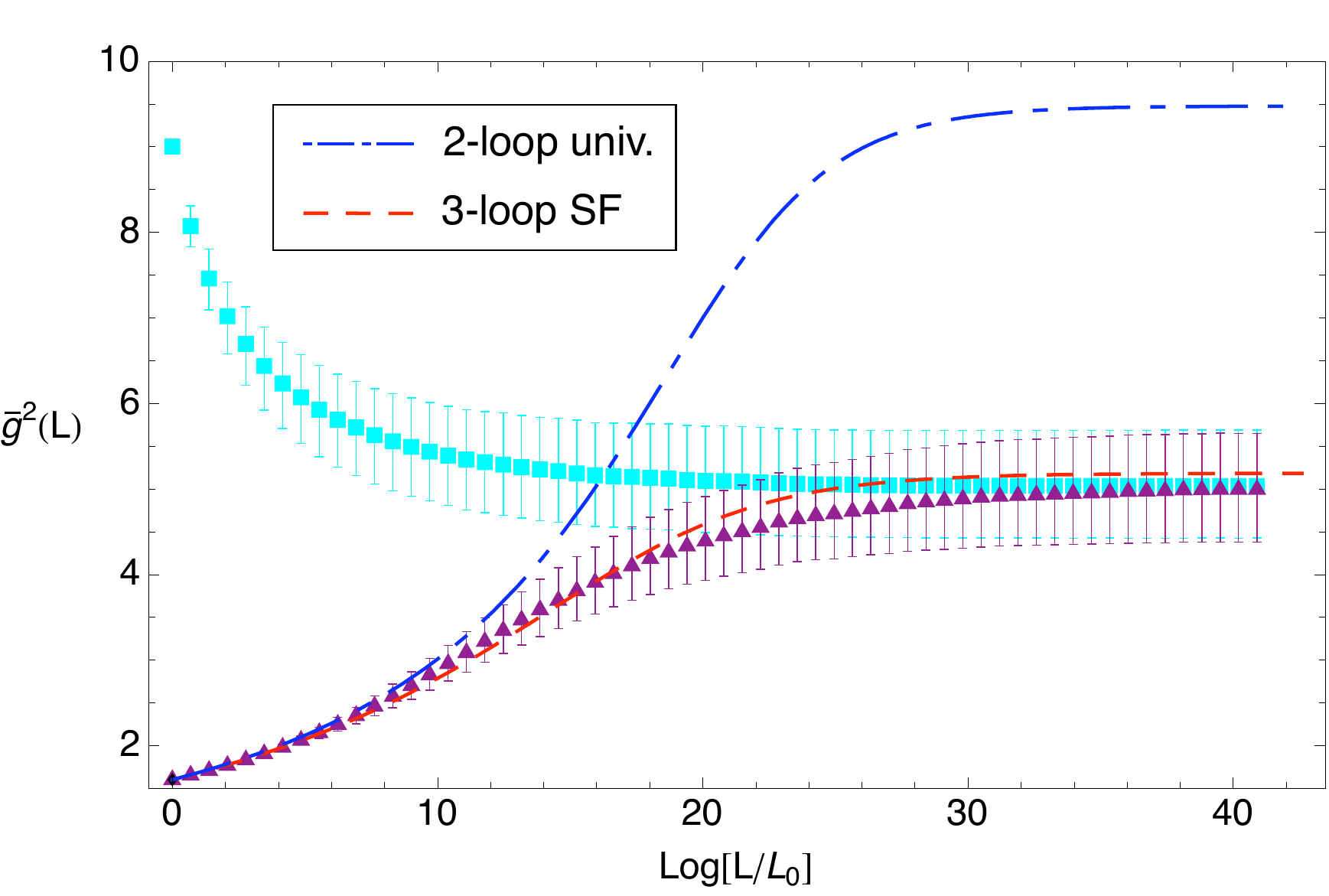}
\caption{\label{fig:ssf-Nf12}Continuum running for $N_f = 12$.  Results shown for running from below the infrared fixed point (purple triangles) are based on $\overline{g}^2(L_0) \equiv 1.6$.  Also shown is continuum backwards running from above the fixed point (light blue squares), based on $\overline{g}^2(L_0) \equiv 9.0$.  Error bars are again purely statistical, although strongly correlated due to the underlying interpolating functions.  Two-loop and three-loop perturbation theory curves are shown for comparison.}
\end{figure}

Our results for the continuum running of $\overline{g}^2(L)$ from below a possible infrared fixed point are shown in
Fig.~\ref{fig:ssf-Nf12}. $L_0$ is again taken to be the scale at which $\overline{g}^2(L_0)
= 1.6$, a relatively weak value. The points are shown for for values
of $L/L_0$ increasing by factors of $2$. The (statistical) errors are derived as described in Appendix~\ref{app:error}. For reference, the two-loop and three-loop perturbative curves for $\overline{g}^2(L)$ are also shown in Fig.~\ref{fig:ssf-Nf12}.  From the figure, we conclude that the infrared behavior is indeed governed by a fixed point whose value lies within the statistical error band.  Because of the underlying use of an interpolating function, the error bars of adjacent points in Fig.~\ref{fig:ssf-Nf12} are highly correlated.  As the running coupling approaches the infrared fixed point, this correlation approaches 100\%, so that the error bars asymptotically approach a stable value as the number of steps is taken to infinity.  The range of possible values of the fixed point from our simulations is consistent with the three-loop perturbative value in the SF scheme. It is well below estimates \cite{Appelquist:1998rb} of the strength required to trigger spontaneous chiral symmetry breaking and confinement.

The infrared fixed point also governs the $L \rightarrow \infty$ behavior starting from
values of $\overline{g}^2(L)$ above the fixed point.  As discussed in Sec.~\ref{sec:latss}, the continuum limit
is then no longer guaranteed to exist and the step-scaling procedure cannot be naively applied.
Instead, one can restrict the discussion to finite but small values of $a/L$, small enough to minimize lattice artifacts but large enough so that for $\overline{g}^2(L)$ near but above the fixed point,
$g_0^{2}(a/L)$ remains small enough not to trigger a bulk phase transition. Since we use a constant extrapolation, this procedure can be taken to define, within our errors,  a $\overline{g}^2(L)$ at a small but finite $a/L$. The step-scaling procedure then leads to the continuum running from above to the fixed point, also shown in Fig.~\ref{fig:ssf-Nf12}. The statistical-error band is derived as in the approach from below.

Finally we note that the exponent $\gamma$ governing the approach to the infrared
fixed point in the SF scheme can also be extracted from the simulation data.  
Taking the log of Eq.~(\ref{eq:anomdim}), we see that the quantity $\log\left[\overline{g}_\star^2 - \overline{g}^2(L)\right]$ should have a linear dependence on $L$ with slope $-\gamma$ near the fixed point.  Computing this quantity from our data, running from either above or below the fixed point, we find $\gamma = 0.13 \pm 0.03$, somewhat smaller than the three-loop SF perturbative estimate of $0.286$.

%%%%%%%%%%%%%%%%%%%%%%%%%%%%%%%%%%%%%%%%%%%%%%%%%%%%%%%%%%%%%%%%%%%%%%%%%%%%%%
\subsection{Comparison with Other Lattice Work}
%%%%%%%%%%%%%%%%%%%%%%%%%%%%%%%%%%%%%%%%%%%%%%%%%%%%%%%%%%%%%%%%%%%%%%%%%%%%%%

%%%%%%%%%%%%%%%%%%%%%%%%%%%%%%%%%%%%%%%%%%%%%%%%%%%%%%%%%%%%%%%%%%%%%%%%%%%%%%
\subsubsection{Schr\"odinger functional studies}
%%%%%%%%%%%%%%%%%%%%%%%%%%%%%%%%%%%%%%%%%%%%%%%%%%%%%%%%%%%%%%%%%%%%%%%%%%%%%%

Lattice simulations for the $\text{SU}(3)$ Schr\"odinger functional running coupling
have been performed for $N_f= 16$ \cite{Heller:1997vh}, for the quenched theory
\cite{Guagnelli:1998ud}, and for $N_f = 2$ \cite{Bode:2001jv}.
For $N_f = 16$ \cite{Heller:1997vh}, the perturbative infrared fixed point is
very weak. In this case, the simulations were done for values of the lattice coupling in
the weak-coupling (chirally symmetric and deconfined) phase but leading to
values of $\overline{g}^2(L)$ well above the perturbative fixed point.
Evidence was presented that $\overline{g}^2(L)$ decreases with increasing
$L$, consistent with the approach to the fixed point from above as expected
with a continuum infrared fixed point. A continuum extrapolation via the
step-scaling procedure was, however, not implemented.

For both the quenched theory \cite{Heitger:2001hs} and $N_f = 2$
\cite{Bode:2001jv}, the step-scaling procedure was implemented and a
continuum running coupling was extracted. In each case, starting with a
$\overline{g}^2(L)$ well into the perturbative regime, the coupling grows
through large, nonperturbative values. And in each case, the growth is more
rapid than for $N_f = 8$, as shown in Fig.~\ref{fig:ssf-Nf8}. For the quenched theory,
$\overline{g}^2(L)$ was argued to grow exponentially at large $L$,
consistent with the leading order prediction from the strong-coupling
expansion \cite{Heitger:2001hs}.

%%%%%%%%%%%%%%%%%%%%%%%%%%%%%%%%%%%%%%%%%%%%%%%%%%%%%%%%%%%%%%%%%%%%%%%%%%%%%%
\subsubsection{Other multifermion studies}
%%%%%%%%%%%%%%%%%%%%%%%%%%%%%%%%%%%%%%%%%%%%%%%%%%%%%%%%%%%%%%%%%%%%%%%%%%%%%%

Lattice simulations of $\text{SU}(3)$ gauge theories with multiple fermions in the fundamental representation began more than 20 years ago. Brown \textit{et al.}~\cite{Brown:1992fz} examined the $N_f = 8$ case with staggered fermions, providing evidence that the theory confines, but remaining inconclusive due to finite volume effects. Damgaard \textit{et al.}~\cite{Damgaard:1997ut} examined the staggered $N_f = 16$ case, noting that even with the
very weak infrared fixed point present in this theory, a bulk chiral transition sets in at sufficiently strong lattice coupling. In a 2001 Columbia PhD thesis \cite{Sui:2001rf}, Sui studied QCD for $N_f = 2$ and $N_f = 4$ staggered fermions, observing stronger finite-lattice-size effects in the latter case. The work of Iwasaki \textit{et al.}~\cite{Iwasaki:2003de} is perhaps most directly relevant to the results reported here and in Ref.~\cite{Appelquist:2007hu}. Through a focus on the strong lattice-coupling phase using Wilson fermions, they concluded that $6 \leq  N_f^\text{c} \leq7$, in disagreement with our results.

Interest in multifermion studies has grown considerably in the past few months.
Deuzeman \textit{et al.}~\cite{Deuzeman:2008sc} have examined chiral symmetry breaking for the $N_f = 8$ case using staggered fermions, concluding that the lower end of the conformal window is indeed above $N_f = 8$. Jin and Mawhinney \cite{Jin:2008qu} have come to the same  conclusion through a study of the chiral condensate and the heavy quark potential. Fodor \textit{et al.}~\cite{Fodor:2008hn} have begun a multifermion simulation using staggered fermions, while Bilgici \textit{et al.}\ have developed a new approach to running coupling measurement with an eye towards eventual multifermion measurements \cite{Bilgici:2008mt}.

%%%%%%%%%%%%%%%%%%%%%%%%%%%%%%%%%%%%%%%%%%%%%%%%%%%%%%%%%%%%%%%%%%%%%%%%%%%%%%
\section{Summary and Discussion}\label{sec:summ}
%%%%%%%%%%%%%%%%%%%%%%%%%%%%%%%%%%%%%%%%%%%%%%%%%%%%%%%%%%%%%%%%%%%%%%%%%%%%%%

We have concluded from lattice simulations of the Schr\"odinger
functional running coupling that for an $\text{SU}(3)$ gauge theory with
$N_f$ Dirac fermions in the fundamental representation, the value $N_f = 8$
lies outside the conformal window, leading to confinement and chiral
symmetry breaking, while $N_f = 12$ lies within the conformal window,
governed by an infrared fixed point. We have bounded the fixed point value
as shown in Fig.~\ref{fig:ssf-Nf12} and estimated the exponent $\gamma$ describing the
approach to the fixed point Eq.~(\ref{eq:anomdim}). This is, as far as we know, the first
nonpertubative evidence for the existence of infrared conformal behavior in
a nonsupersymmetric gauge theory. These results confirm and
refine the analysis of Ref.~\cite{Appelquist:2007hu}.

The $N_f = 8$ and $N_f = 12$ results imply that the lower end of the
conformal window, $N_f^\text{c}$, lies in the range $ 8 < N_f^\text{c} <
12$. This conclusion, in disagreement with Ref.~\cite{Iwasaki:2003de}, is
reached employing the Schr\"odinger functional (SF) running coupling,
$\overline{g}^2(L)$, defined {\it at the box boundary $L$} with a set of
special boundary conditions. This coupling is a gauge invariant quantity, valid
for any coupling strength and running in accordance with perturbation theory
at short distances.

For $N_f$=$8$, we have simulated $\overline{g}^2(L)$ up through values that
exceed typical estimates of the coupling strength required to trigger
dynamical chiral symmetry breaking \cite{Appelquist:1998rb}, with no
evidence for an infrared fixed point or even an inflection point.  For $N_f$=$12$, our observed infrared fixed point is rather weak, agreeing within the estimated errors with the three-loop fixed point in the SF scheme, and well below typical estimates of the coupling strength required to trigger
dynamical chiral symmetry breaking \cite{Appelquist:1998rb}.

Whether perturbation theory can be used reliably to reproduce the behavior in the
vicinity of the $N_{f} = 12$ fixed point remains to be seen. The three-loop
value of the fixed point is substantially different from the two-loop value.
On the other hand, in the $\overline{\text{MS}}$ scheme where the four-loop
beta function has been computed, the four-loop fixed point is shifted by
only a small amount from the three-loop value. The relative weakness of this fixed point, together with the fact that $N_f^\text{c}$ cannot be
much smaller, raises the question of whether the theory remains perturbative
throughout the conformal window as suggested by Gardi and Grunberg \cite{Gardi:1998ch}.
If this is the case, the behavior in the neighborhood of $N_f^\text{c}$ would be rather
different from the supersymmetric $\text{SU}(N)$ gauge theory \cite{Seiberg:1994pq}. In particular,
it is not clear whether there would be a useful, effective low energy description
of the infrared behavior.

It is important to confirm our results by employing other definitions
of the running coupling, for example, based on the Wilson loop and static potential \cite{Bilgici:2008mt}, and by examining scheme-independent
quantities. Most notably, spontaneously chiral symmetry breaking as a function of $N_f$ must be studied through a zero-temperature
lattice simulation of the chiral condensate \cite{Jin:2008qu}.  Simulations of $\overline{g}^2(L)$ for other values of
$N_f$, in particular $N_f$=$10$, are crucial to determine more
accurately the lower end of the conformal window and to study the phase
transition as a function of $N_f$.
All of these analyses should be extended to other gauge groups and other
representation assignments for the fermions \cite{Svetitsky:2008bw,DeGrand:2008dh,Shamir:2008pb,Catterall:2007yx,Catterall:2008qk,DelDebbio:2008wb,Hietanen:2008vc,Hietanen:2008mr,DelDebbio:2008tv,Fodor:2008hm,DelDebbio:2008zf}.

The phenomenological relevance of these studies remains to be seen. One possibility is that a theory with infrared conformal symmetry could describe some new sector, coupled to the standard model through gauge-singlet operators \cite{Georgi:2007ek}. Another possibility, much discussed in the literature, is that a theory with $N_f$ outside but near the conformal window ($\lesssim N_f^\text{c}$) could describe electroweak breaking and provide the basis for walking technicolor \cite{Holdom:1981rm, Yamawaki:1985zg, Appelquist:1987fc}. In this class of theories, as $N_f \rightarrow N_f^\text{c}$ from below, a hierarchy emerges between the electroweak scale and the larger mass scale where the gauge coupling becomes strong. This could be signaled by the appearance of a plateau of finite extent in  $\overline{g}^2(L)$, and by the development of a hierarchy between the chiral condensate and the electroweak scale. It is also important to explore the particle spectrum in this limit and to compute the electroweak precision parameters, in particular the $S$ parameter \cite{Peskin:1991sw}. These studies are currently
underway \cite{Fleming:2008jx}.

%%%%%%%%%%%%%%%%%%%%%%%%%%%%%%%%%%%%%%%%%%%%%%%%%%%%%%%%%%%%%%%%%%%%%%%%%%%%%%
\begin{acknowledgments}
%%%%%%%%%%%%%%%%%%%%%%%%%%%%%%%%%%%%%%%%%%%%%%%%%%%%%%%%%%%%%%%%%%%%%%%%%%%%%%

We acknowledge helpful discussions with Urs Heller, Walter Goldberger, Pavlos Vranas, and Rich
Brower.  This work was supported partially
by DOE Grant No.~DE-FG02-92ER-40704 (T.A.\ and E.N.), by the National Science Foundation through TeraGrid resources \cite{Teragrid} provided by the Texas Advanced Computing Center, and took place partly at the Aspen Center for Physics (TA).  Simulations were based in part on the MILC code \cite{MILC}, and were performed at Fermilab and Jefferson Lab on clusters provided by the DOE's
U.S.~Lattice QCD (USQCD) program, the Ranger cluster at the Texas Advanced Computing Center supported under TeraGrid allocation TG-MCA08X008, the Yale Life Sciences Computing Center
supported under NIH Grant No.~RR 19895, the Yale High Performance Computing
Center, and on a SiCortex SC648 from SiCortex, Inc.  

%%%%%%%%%%%%%%%%%%%%%%%%%%%%%%%%%%%%%%%%%%%%%%%%%%%%%%%%%%%%%%%%%%%%%%%%%%%%%%
\end{acknowledgments}
%%%%%%%%%%%%%%%%%%%%%%%%%%%%%%%%%%%%%%%%%%%%%%%%%%%%%%%%%%%%%%%%%%%%%%%%%%%%%%

\appendix

%%%%%%%%%%%%%%%%%%%%%%%%%%%%%%%%%%%%%%%%%%%%%%%%%%%%%%%%%%%%%%%%%%%%%%%%%%%%%%
\section{Statistical and systematic error in step-scaling analysis}\label{app:error}
%%%%%%%%%%%%%%%%%%%%%%%%%%%%%%%%%%%%%%%%%%%%%%%%%%%%%%%%%%%%%%%%%%%%%%%%%%%%%%

%%%%%%%%%%%%%%%%%%%%%%%%%%%%%%%%%%%%%%%%%%%%%%%%%%%%%%%%%%%%%%%%%%%%%%%%%%%%%%
\subsection{Numerical-simulation error}
%%%%%%%%%%%%%%%%%%%%%%%%%%%%%%%%%%%%%%%%%%%%%%%%%%%%%%%%%%%%%%%%%%%%%%%%%%%%%%

  The hybrid molecular dynamics (HMD) method involves the solution of classical equations of motion,
  requiring a numerical integration at a finite step size
  $\Delta \tau$.  This introduces systematic numerical errors of $O((\Delta \tau)^2)$
  into all observables, including the running coupling.  Removal of this finite step-size error from a given measurement of $\overline{g}^2(\beta, L)$ can be accomplished by simulating at multiple values of $\Delta \tau$ and performing a quadratic extrapolation to zero.  At relatively weak values of the bare coupling, the step-size error is observed to reduce the measured values of $\overline{g}^2$.  Furthermore, the magnitude of the shift increases with the box size $L/a$.  Although statistically significant, the effect on step-scaling results is negligible for $L/a = 8$ and below, but becomes significant at larger box sizes.  At stronger values of the bare coupling, no systematic shift due to step-size error can be resolved within our statistical error.  Wherever the effect or step-size error is observed to be significant, only extrapolated values of $\overline{g}^2$ are used in our analysis, effectively removing this source of systematic error.
  
  Although our goal is to generate a set
  of statistically independent gauge configurations, in practice
  configurations which are separated by only a small number of updates
  (by a small MD time) can be correlated with each other.  The
  existence of these ``autocorrelations" can lead to underestimation of
  statistical errors if classical estimators are used.  Even with appropriate
  statistical methods, the time series must extend to several times the
  autocorrelation time of the observable in order to obtain an unbiased
  measurement. Statistical estimates of the integrated autocorrelation time show a clear trend towards longer autocorrelations as either the box size $L/a$ or the strength of the lattice coupling $g_0^2$ is increased, with the longest estimated time on the order of $1 000$ trajectories.

In addition, autocorrelation times in certain time series are, in effect, enhanced by the observed phenomenon of ``excursions."  This has been noted in prior studies of the SF running coupling \cite{Luscher:1993gh, DellaMorte:2004bc}.  The coupling is seen to jump to a new equilibrium value, which can remain stable for up to several thousand trajectories.  This can be interpreted as a tunneling of the system from near the original minimum of the action, determined by the imposed background field, into other metastable minima.  In the presence of these excursions, autocorrelations in the observable are introduced on the scales of the average duration and period of the tunneling events.

Although the excursions themselves have a physical interpretation, the correlations that they induce in the data are artifacts of the procedure used to generate gauge configurations, and the associated time scales are dependent on the choice of such a procedure.  For example, an update algorithm based on the selection of completely random gauge configurations would have no autocorrelations by design, but could still show evidence of tunneling into secondary minima in the form of non-Gaussianity in a histogram of measured coupling strengths.

\begin{figure}
\includegraphics[width=160mm]{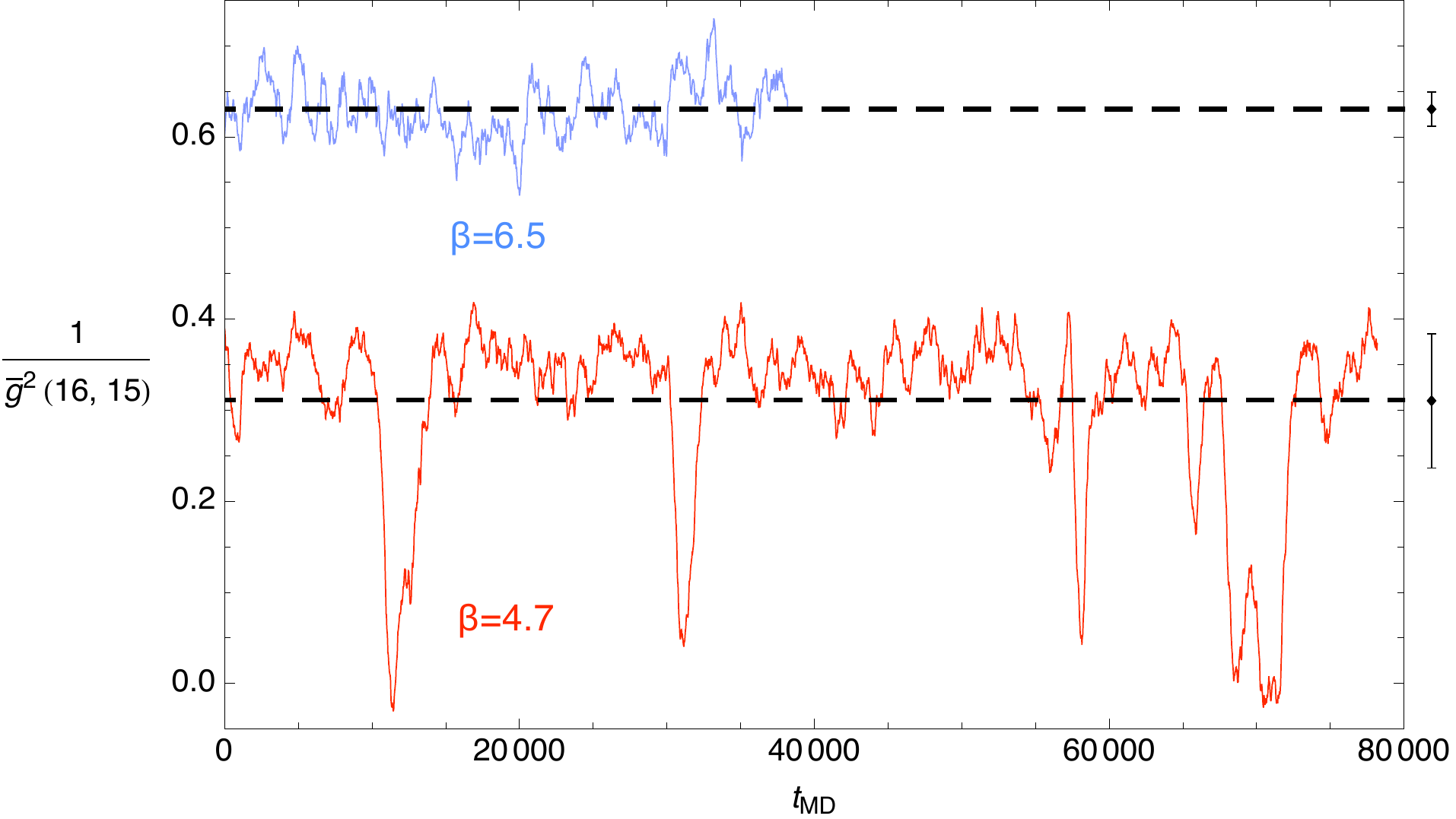}
\caption{\label{fig:excursion}Representative time series taken from our
12-flavor simulations.  The estimated mean values with statistical error for each time series are depicted in the dotted lines and the points with error bars to the right of the plot.  The time-series plot is a running average over a window of 800 points.  Both data sets were gathered on volumes of $16^3 \times 15$, and represent relatively weak coupling ($\beta = 6.5$, light blue)
and somewhat stronger coupling ($\beta = 4.7$, dark red).  The difference in length between the two time-series reflects our choice of target numbers of trajectories, as discussed in the text.}
\end{figure}

The excursions are empirically seen to occur always to an equilibrium at stronger
coupling $\overline{g}^2(L)$, and become more frequent as the bare coupling strength is increased
from weak to stronger values.  In particular we observe tunneling events in the running average of the time series for $5.8 \leq \beta \leq 4.7$, at both eight and twelve flavors, with these events becoming impossible to isolate from statistical fluctuations at stronger coupling.
Some representative plots showing this effect are shown in Fig.~\ref{fig:excursion}.
The contrast between the two time series is sharp, with tunneling events
clearly evident only in the stronger-coupling time series.

We therefore choose the target number of trajectories for a particular measurement of $\overline{g}^2(\beta, L)$ as follows: $20 000$ trajectories for $\beta \geq 8.0$; $40 000$ for $5.0 \geq \beta > 8.0$; and $80 000$ for $\beta < 5.0$. These values exceed all above estimates of the autocorrelation time,
and allow sampling of multiple excursions where such events are observed. 

Estimation of statistical error, with full propagation of errors including continuum-extrapolation error through all step-scaling steps, is performed using the bootstrap method.  The raw data are first reduced to uncorrelated blocks.  Two thousand bootstrap replications of the data set are generated, and quantities of interest are computed as statistics on the bootstrap data.  Two-sided errors are shown in all cases, representing $1\sigma$ confidence intervals on the mean, computed using the bias-corrected and accelerated confidence interval estimation method \cite{Efron:1993ok}.  Measured values for $\overline{g}^2(L)$ with estimated two-sided error bars are included in Tables~\ref{table:Nf8} and \ref{table:Nf12}.  Although we quote a single value for each $(\beta, L/a)$ in the data tables, we emphasize that our analysis is carried through on the complete sets of bootstrap replicated data from which these mean values are derived.  

%%%%%%%%%%%%%%%%%%%%%%%%%%%%%%%%%%%%%%%%%%%%%%%%%%%%%%%%%%%%%%%%%%%%%%%%%%%%%%
\subsection{Interpolating-function error}
%%%%%%%%%%%%%%%%%%%%%%%%%%%%%%%%%%%%%%%%%%%%%%%%%%%%%%%%%%%%%%%%%%%%%%%%%%%%%%

The choice of a particular interpolating functional form is a potential source of systematic error, particularly if it yields a poor fit to the data, reflected by a value of $\chi^2$ per degree of freedom (DOF) significantly larger than 1.  Using our interpolating functional form Eq.~(\ref{eq:interpolating_function}), we find an excellent fit to the simulation data; at $N_f = 8$, we find an overall \chidof $= 1.62 \pm 0.30$ with 107 degrees of freedom, while for the $N_f = 12$ data the fit yields \chidof $= 1.47 \pm 0.26$ with 171 degrees of freedom.  Errors on $\chi^2$ are estimated using the jackknife method.  Note that these are global $\chi^2$ values, representing a sum over contributions from all $L/a$ fits, divided by the total number of degrees of freedom.  The relatively low probability of these values in the $\chi^2$ distribution indicates that our error bars are likely somewhat underestimated.  Possible sources of this effect are detailed in the previous subsection.

We have further attempted to search for systematic error due to the interpolating function by trying to fit a variety of other functional forms.  One approach is to attempt to improve upon the form in Eq.~(\ref{eq:interpolating_function}) by the addition of extra terms.  We have considered the addition of terms nonanalytic in $g_0^2$, and the addition of terms and constraints which reproduce one-loop perturbation theory in the limit $g_0^2 \rightarrow 0$.  In each case, the additional terms do not significantly improve \chidof, and the results of the analysis based on such fits are indistinguishable from those based on Eq.(\ref{eq:interpolating_function}).

Another possibility is to use an altogether different functional form, such as the Laurent series expansion of Ref.~\cite{Appelquist:2007hu}.  These forms can also be extended with nonanalytic terms and perturbative constraints.  For $N_f = 12$, large systematic shifts are seen in the continuum running curves based on such fits.  This systematic effect is universally associated with a significantly higher \chidof, indicating that the interpolating function does not accurately reflect the underlying measurements.

We stress that since our fit functions are used only for interpolation, not extrapolation, any two fits to the coupling measurements which yield comparable and acceptably small \chidof\ will give indistinguishable results for the continuum running.

We conclude that the systematic error associated with the selection of the final form Eq.~(\ref{eq:interpolating_function}) is negligible, given the values of \chidof\ quoted above.

%%%%%%%%%%%%%%%%%%%%%%%%%%%%%%%%%%%%%%%%%%%%%%%%%%%%%%%%%%%%%%%%%%%%%%%%%%%%%%
\subsection{Continuum-extrapolation error} \label{sec:sserror}
%%%%%%%%%%%%%%%%%%%%%%%%%%%%%%%%%%%%%%%%%%%%%%%%%%%%%%%%%%%%%%%%%%%%%%%%%%%%%%

Since each step in the step-scaling procedure involves a
continuum extrapolation of the step-scaling function $\Sigma(2,u,a/L)$,
the choice of functional form in the extrapolation is another potential source of systematic error.

What is the expected behavior of $\Sigma(2,u,a/L)$ as a function of $a/L$?  Staggered fermions with the Wilson gauge action inherently
lead to discretization effects of $O(a^2)$.  However, the presence of
the Dirichlet boundaries here leads to operators which contribute $O(a)$
lattice artifacts, as shown in Eq.~(\ref{gaugeact}).  We include a counterterm for
these operators, with its value determined by one-loop perturbation
theory. With this counterterm, we expect
that the $O(a)$ terms are small, even more so in $\Sigma(2,u,a/L)$ where much of the $a/L$ dependence is removed.

By comparing our data to perturbative running at fixed, relatively weak lattice coupling, we find that in fact all lattice-artifact corrections are negligible compared to our statistical errors for $L/a \geq 8$.
Independent of perturbation theory, we find that the best fit (smallest \chidof) for the continuum extrapolation at $N_f = 12$ is given by a constant, $a/L$-independent extrapolation, yielding $\sigma(2,u)$ as a weighted average of $\Sigma(2,u,a/L)$ over the available range of $a/L$.  Based on this experience, we use constant extrapolation at $N_f = 8$ as well. Lattice artifacts play a less important role here because of the stronger continuum running (which also provides justification for not including data at $L/a = 20$). The
absence of $L/a = 20$ data at $N_f = 8$ means that there are only two available steps ($6 \rightarrow 12$ and $8 \rightarrow 16$). Thus the constant $a/L$ extrapolation is the only constrained fit available. The errors for each $\Sigma(2,u,a/L)$ lead to statistical errors in $\sigma(2,u)$, which is represented by a statistical-error band for the final continuum-running curve.

Statistical error in the continuum extrapolation is computed by the bootstrap method; the extrapolation is performed independently on each bootstrap ensemble, leading to a distribution of values for $\sigma(2,u)$, which can then be used to estimate a mean value and two-sided confidence interval.  The result of the application of $n$ steps, $\sigma(2^n, u)$, is likewise computed within each bootstrap ensemble to obtain a full distribution.

%%%%%%%%%%%%%%%%%%%%%%%%%%%%%%%%%%%%%%%%%%%%%%%%%%%%%%%%%%%%%%%%%%%%%%%%%%%%%%
\section{Running coupling data}\label{app:data}
%%%%%%%%%%%%%%%%%%%%%%%%%%%%%%%%%%%%%%%%%%%%%%%%%%%%%%%%%%%%%%%%%%%%%%%%%%%%%%
Our measurements of the running coupling $\overline{g}^2(L)$ are presented in Tables \ref{table:Nf8} and \ref{table:Nf12} below.  Two-sided error bars are estimated using the BCa method as described in Appendix~\ref{app:error}.

\pagebreak
\begin{center}
\begin{longtable}{|c|ccccc|}
\caption[$N_f = 8$ data]{Measurements of $\overline{g}^2(\beta, L/a)$, $N_f = 8$.} \label{table:Nf8}\\

\hline \hline
$\bar{g}^2(L)$&&&$L/a$&&\\
\hline
$\beta$&6&8&10&12&16\\
\hline
\endfirsthead

\hline
$\bar{g}^2(L)$&&&$L/a$&&\\
\hline
$\beta$&6&8&10&12&16\\
\hline
\endhead

\hline
\endfoot

\hline \hline
\endlastfoot
4.550&13.06\errpm{+34}{-33}&17.12\errpm{+61}{-64}&&&\\
4.560&12.10\errpm{+32}{-32}&&&&\\
4.570&11.51\errpm{+20}{-24}&&&&\\
4.580&10.21\errpm{+19}{-19}&&&&\\
4.590&10.00\errpm{+22}{-22}&&&&\\
4.600&9.78\errpm{+24}{-24}&11.6\errpm{+1.2}{-1.3}&14.40\errpm{+69}{-69}&21.9\errpm{+10.0}{-9.4}&\\
4.650&7.887\errpm{+99}{-95}&9.63\errpm{+26}{-26}&11.23\errpm{+34}{-34}&11.5\errpm{+1.1}{-1.0}&\\
4.700&6.91\errpm{+14}{-12}&8.21\errpm{+50}{-55}&10.85\errpm{+68}{-66}&&\\
4.800&5.626\errpm{+98}{-79}&6.56\errpm{+25}{-25}&7.11\errpm{+23}{-22}&7.61\errpm{+41}{-35}&9.5\errpm{+1.4}{-1.2}\\
4.900&4.761\errpm{+51}{-52}&5.18\errpm{+16}{-17}&&&\\
5.000&4.204\errpm{+54}{-65}&4.68\errpm{+15}{-15}&4.96\errpm{+17}{-12}&5.64\errpm{+34}{-31}&6.45\errpm{+57}{-49}\\
5.100&3.788\errpm{+43}{-39}&4.2\errpm{+0.1}{-0.1}&&&\\
5.200&3.382\errpm{+42}{-38}&3.674\errpm{+63}{-61}&&&\\
5.300&3.087\errpm{+26}{-24}&3.311\errpm{+35}{-32}&&&\\
5.400&2.972\errpm{+29}{-29}&3.145\errpm{+45}{-37}&&&\\
5.500&2.724\errpm{+22}{-22}&2.965\errpm{+39}{-33}&3.122\errpm{+71}{-48}&3.380\errpm{+90}{-76}&3.374\errpm{+55}{-65}\\
5.600&2.603\errpm{+11}{-12}&2.795\errpm{+43}{-31}&&&\\
5.700&2.4424\errpm{+100}{-93}&2.590\errpm{+13}{-13}&&&\\
5.800&2.3340\errpm{+82}{-86}&2.491\errpm{+17}{-13}&&&\\
5.830&2.286\errpm{+9}{-11}&2.456\errpm{+15}{-15}&2.535\errpm{+12}{-13}&2.647\errpm{+26}{-23}&2.842\errpm{+41}{-44}\\
5.900&2.2246\errpm{+90}{-82}&2.340\errpm{+12}{-12}&&&\\
6.000&2.1374\errpm{+56}{-58}&2.264\errpm{+12}{-11}&&&\\
6.100&2.0578\errpm{+63}{-64}&2.1622\errpm{+98}{-91}&&&\\
6.200&1.9778\errpm{+47}{-46}&2.0687\errpm{+84}{-94}&&&\\
6.300&1.9127\errpm{+58}{-52}&2.0008\errpm{+88}{-85}&&&\\
6.400&1.8430\errpm{+93}{-55}&1.9230\errpm{+55}{-60}&&&\\
6.500&1.7869\errpm{+50}{-44}&1.8601\errpm{+73}{-76}&&&\\
6.590&1.7328\errpm{+36}{-31}&&&&\\
6.600&1.7232\errpm{+39}{-33}&1.7996\errpm{+44}{-46}&&&\\
6.700&1.6665\errpm{+36}{-44}&1.7347\errpm{+57}{-63}&&&\\
6.800&1.6253\errpm{+25}{-26}&1.6938\errpm{+53}{-56}&1.7359\errpm{+82}{-68}&&\\
6.883&1.5865\errpm{+24}{-21}&&&1.7143\errpm{+82}{-73}&\\
6.900&1.5776\errpm{+33}{-31}&1.6260\errpm{+51}{-51}&&&\\
7.000&1.5323\errpm{+34}{-38}&1.5845\errpm{+33}{-33}&&&1.734\errpm{+9}{-11}\\
7.090&&&1.585\errpm{+14}{-13}&&\\
7.100&1.4886\errpm{+22}{-23}&&1.5882\errpm{+60}{-62}&&\\
7.115&&&1.5787\errpm{+77}{-80}&&\\
7.153&&&1.5664\errpm{+58}{-65}&&\\
7.200&1.4559\errpm{+28}{-27}&&&&\\
7.300&1.4133\errpm{+27}{-29}&&&&\\
7.400&1.3795\errpm{+23}{-27}&&&&\\
7.500&1.3436\errpm{+24}{-23}&1.3880\errpm{+31}{-32}&&&\\
8.000&1.2000\errpm{+18}{-19}&1.2302\errpm{+35}{-35}&&&\\
8.500&1.0863\errpm{+17}{-18}&1.1141\errpm{+24}{-24}&&&\\
9.000&0.99470\errpm{+98}{-95}&1.0189\errpm{+29}{-27}&&&\\
12.000&0.65805\errpm{+44}{-43}&0.66857\errpm{+77}{-84}&&0.6834\errpm{+47}{-51}&0.7054\errpm{+68}{-68}\\
12.800&0.6036\errpm{+9}{-10}&0.61228\errpm{+98}{-99}&&&\\
13.710&0.55193\errpm{+67}{-60}&0.5618\errpm{+11}{-11}&&&\\
14.770&0.50383\errpm{+49}{-48}&0.50818\errpm{+73}{-84}&&&\\
16.000&0.45600\errpm{+46}{-46}&0.46117\errpm{+65}{-59}&&&\\
17.450&0.41039\errpm{+33}{-34}&0.41453\errpm{+51}{-57}&&&\\
19.200&0.36557\errpm{+28}{-31}&0.36869\errpm{+34}{-38}&&&\\
21.330&0.32349\errpm{+20}{-21}&0.32680\errpm{+30}{-33}&&&\\
24.000&0.282902\errpm{+97}{-92}&0.28465\errpm{+17}{-18}&&0.28876\errpm{+88}{-78}&\\
27.430&0.24331\errpm{+20}{-19}&0.24490\errpm{+26}{-25}&&&\\
32.000&0.20540\errpm{+18}{-17}&0.20632\errpm{+22}{-22}&&&\\
38.400&0.16861\errpm{+16}{-15}&0.16893\errpm{+23}{-26}&&&\\
48.000&0.13248\errpm{+9}{-10}&0.133007\errpm{+99}{-94}&&0.13361\errpm{+43}{-39}&\\
64.000&0.097919\errpm{+42}{-46}&0.098113\errpm{+76}{-86}&&&\\
96.000&0.064344\errpm{+31}{-34}&0.064433\errpm{+21}{-18}&&&\\
192.000&0.031717\errpm{+18}{-18}&0.031675\errpm{+26}{-28}&&&\\
\end{longtable}
\end{center}

\begin{center}
\begin{longtable}{|c|cccc|}
\caption[$N_f = 8$ pars]{Interpolation best-fit parameters, $N_f = 8$.} \label{table:Nf8pms}\\

\hline
&&$L/a$&&\\
\hline
param&6&8&12&16\\
\hline
\endhead

\hline
$\chi^2$/dof&1.78&1.49&1.33&1.65\\
$N_{dof}$&55&43&7&2\\
\hline
\endfoot
$c_{1,L/a}$&0.4632(99)&0.4932(59)&0.581(18)&1.01(18)\\
$c_{2,L/a}$&-0.14(12)&-0.167(44)&-0.235(40)&-1.01(37)\\
$c_{3,L/a}$&1.13(59)&0.76(11)&0.245(23)&0.60(19)\\
$c_{4,L/a}$&-3.2(1.3)&-0.98(11)&0&0\\
$c_{5,L/a}$&4.7(1.6)&0.441(37)&0&0\\
$c_{6,L/a}$&-3.43(92)&0&0&0\\
$c_{7,L/a}$&0.98(21)&0&0&0\\
\end{longtable}
\end{center}

\pagebreak

\begin{center}
\begin{longtable}{|c|cccccc|}
\caption[$N_f = 12$ data]{Measurements of $\overline{g}^2(\beta, L/a)$, $N_f = 12$.} \label{table:Nf12}\\

\hline \hline
$\bar{g}^2(L)$&&&$L/a$&&&\\
\hline
$\beta$&6&8&10&12&16&20\\
\hline
\endfirsthead

\hline
$\bar{g}^2(L)$&&&$L/a$&&&\\
\hline
$\beta$&6&8&10&12&16&20\\
\hline
\endhead

\hline
\endfoot

\hline \hline
\endlastfoot
\hline
4.200&14.84\errpm{+26}{-26}&11.31\errpm{+56}{-53}&13.0\errpm{+1.7}{-1.6}&12.5\errpm{+1.1}{-1.1}&&\\
4.210&13.63\errpm{+27}{-25}&11.12\errpm{+93}{-87}&&&&\\
4.220&12.44\errpm{+21}{-23}&11.1\errpm{+0.8}{-0.8}&&&&\\
4.230&11.6\errpm{+0.2}{-0.2}&8.9\errpm{+1.6}{-1.7}&&&&\\
4.240&11.10\errpm{+17}{-17}&7.7\errpm{+1.5}{-1.3}&&&&\\
4.250&10.48\errpm{+24}{-25}&9.14\errpm{+37}{-35}&9.32\errpm{+80}{-76}&&&\\
4.270&9.47\errpm{+13}{-11}&8.75\errpm{+42}{-41}&&&&\\
4.300&8.3\errpm{+0.1}{-0.1}&7.42\errpm{+30}{-29}&7.97\errpm{+37}{-33}&7.49\errpm{+41}{-43}&8.38\errpm{+99}{-96}&6.62\errpm{+49}{-48}\\
4.350&6.91\errpm{+14}{-13}&6.84\errpm{+39}{-38}&6.84\errpm{+27}{-27}&7.39\errpm{+39}{-38}&&\\
4.400&6.140\errpm{+87}{-77}&6.08\errpm{+30}{-29}&6.14\errpm{+36}{-38}&5.95\errpm{+53}{-48}&&\\
4.450&5.474\errpm{+57}{-60}&5.37\errpm{+23}{-24}&6.00\errpm{+23}{-27}&&&\\
4.500&5.151\errpm{+86}{-86}&5.05\errpm{+12}{-12}&5.26\errpm{+13}{-15}&4.97\errpm{+22}{-16}&&\\
4.600&4.248\errpm{+47}{-41}&4.46\errpm{+15}{-13}&4.42\errpm{+14}{-13}&4.22\errpm{+17}{-15}&4.01\errpm{+23}{-19}&\\
4.700&3.822\errpm{+53}{-49}&3.746\errpm{+53}{-48}&3.669\errpm{+120}{-74}&3.81\errpm{+29}{-27}&5.00\errpm{+53}{-48}&\\
4.800&3.458\errpm{+33}{-30}&3.549\errpm{+96}{-96}&3.58\errpm{+12}{-11}&&&\\
4.900&3.061\errpm{+43}{-37}&3.281\errpm{+85}{-77}&3.196\errpm{+57}{-51}&&&\\
5.000&2.884\errpm{+25}{-22}&2.912\errpm{+40}{-28}&3.005\errpm{+85}{-66}&3.023\errpm{+84}{-71}&3.28\errpm{+26}{-25}&\\
5.100&2.733\errpm{+43}{-30}&2.852\errpm{+77}{-77}&2.811\errpm{+24}{-23}&&&\\
5.200&2.549\errpm{+13}{-13}&2.573\errpm{+28}{-22}&&&&\\
5.300&2.466\errpm{+30}{-25}&2.438\errpm{+43}{-26}&2.528\errpm{+27}{-28}&&&\\
5.400&2.325\errpm{+25}{-18}&2.337\errpm{+16}{-17}&&&&\\
5.500&2.1985\errpm{+120}{-97}&2.2346\errpm{+77}{-85}&2.273\errpm{+12}{-12}&2.271\errpm{+27}{-28}&2.360\errpm{+31}{-29}&2.311\errpm{+54}{-62}\\
5.600&2.0979\errpm{+45}{-44}&2.148\errpm{+13}{-15}&&&&\\
5.700&2.0139\errpm{+48}{-55}&2.066\errpm{+11}{-10}&2.1143\errpm{+65}{-77}&&&\\
5.800&1.9461\errpm{+71}{-59}&2.016\errpm{+31}{-24}&&&&\\
5.900&1.8636\errpm{+35}{-36}&1.8970\errpm{+87}{-86}&1.935\errpm{+12}{-13}&&&\\
6.000&1.8039\errpm{+48}{-44}&1.8218\errpm{+40}{-34}&1.879\errpm{+11}{-11}&1.873\errpm{+17}{-16}&1.912\errpm{+18}{-18}&1.922\errpm{+22}{-20}\\
6.100&1.7532\errpm{+27}{-26}&1.7862\errpm{+62}{-61}&1.813\errpm{+11}{-11}&&&\\
6.200&1.6909\errpm{+48}{-44}&1.7400\errpm{+85}{-80}&&&&\\
6.300&1.6405\errpm{+20}{-19}&1.6698\errpm{+63}{-61}&1.6901\errpm{+69}{-80}&&&\\
6.400&1.5827\errpm{+24}{-20}&1.6299\errpm{+52}{-58}&&&&\\
6.500&1.5459\errpm{+24}{-24}&1.5702\errpm{+52}{-51}&1.5967\errpm{+89}{-83}&1.600\errpm{+17}{-16}&1.614\errpm{+13}{-13}&\\
6.600&1.5031\errpm{+16}{-15}&1.5274\errpm{+27}{-25}&&&&\\
6.800&1.4253\errpm{+13}{-14}&1.4509\errpm{+30}{-31}&&&&\\
7.000&1.3471\errpm{+27}{-24}&1.407\errpm{+25}{-23}&1.3985\errpm{+52}{-48}&1.435\errpm{+19}{-18}&&\\
7.200&1.2932\errpm{+12}{-12}&1.3106\errpm{+23}{-23}&&&&\\
7.400&1.2354\errpm{+11}{-11}&1.2538\errpm{+23}{-22}&&&&\\
7.500&&1.2163\errpm{+84}{-81}&&1.2440\errpm{+86}{-85}&1.2465\errpm{+70}{-65}&\\
7.600&1.18292\errpm{+120}{-95}&1.2000\errpm{+18}{-17}&&&&\\
7.800&1.13538\errpm{+80}{-85}&1.1516\errpm{+18}{-19}&&&&\\
8.000&1.0880\errpm{+16}{-14}&1.1027\errpm{+64}{-58}&1.1197\errpm{+33}{-30}&1.1332\errpm{+120}{-92}&1.1312\errpm{+89}{-94}&1.1415\errpm{+80}{-67}\\
12.000&0.62354\errpm{+61}{-63}&0.62833\errpm{+94}{-80}&0.63109\errpm{+87}{-91}&&&\\
12.800&0.57479\errpm{+45}{-48}&0.5798\errpm{+9}{-10}&0.58266\errpm{+91}{-79}&&&\\
13.710&0.52879\errpm{+53}{-49}&0.53198\errpm{+62}{-70}&0.53620\errpm{+75}{-82}&&&\\
14.770&0.48276\errpm{+38}{-34}&0.48656\errpm{+64}{-61}&0.48839\errpm{+52}{-57}&&&\\
16.000&0.43917\errpm{+31}{-30}&0.44166\errpm{+53}{-49}&0.44383\errpm{+84}{-84}&&0.4482\errpm{+19}{-17}&0.4479\errpm{+26}{-25}\\
17.450&0.39659\errpm{+32}{-31}&0.39940\errpm{+35}{-35}&0.40084\errpm{+60}{-62}&&&\\
19.200&0.35498\errpm{+23}{-25}&0.35677\errpm{+32}{-31}&0.35844\errpm{+45}{-38}&&&\\
21.330&0.31519\errpm{+18}{-18}&0.31732\errpm{+27}{-29}&0.31837\errpm{+49}{-49}&&&\\
24.000&0.27629\errpm{+14}{-14}&0.27779\errpm{+23}{-23}&0.27897\errpm{+45}{-40}&0.27852\errpm{+21}{-24}&&\\
27.430&0.23839\errpm{+16}{-14}&0.23937\errpm{+14}{-15}&0.24030\errpm{+30}{-26}&0.23987\errpm{+24}{-23}&&\\
32.000&0.20157\errpm{+13}{-13}&0.20222\errpm{+18}{-18}&0.20323\errpm{+29}{-25}&0.20257\errpm{+22}{-22}&&\\
38.400&0.16603\errpm{+14}{-14}&0.16643\errpm{+20}{-18}&0.16710\errpm{+26}{-25}&0.16740\errpm{+18}{-16}&&\\
48.000&0.13127\errpm{+13}{-13}&0.13124\errpm{+14}{-15}&0.13178\errpm{+19}{-22}&0.13146\errpm{+14}{-14}&&\\
64.000&0.097222\errpm{+42}{-46}&0.097401\errpm{+100}{-82}&0.097317\errpm{+68}{-64}&0.097257\errpm{+84}{-84}&&\\
96.000&0.063998\errpm{+58}{-53}&0.064060\errpm{+28}{-28}&0.063967\errpm{+85}{-81}&0.064089\errpm{+46}{-51}&&\\
192.000&0.031638\errpm{+24}{-22}&0.031645\errpm{+25}{-24}&0.031677\errpm{+18}{-18}&0.031673\errpm{+11}{-12}&0.031675\errpm{+34}{-32}&\\
\hline
\end{longtable}
\end{center}

\begin{center}
\begin{longtable}{|c|cccccc|}
\caption[$N_f = 12$ pars]{Interpolation best-fit parameters, $N_f = 12$.} \label{table:Nf12pms}\\

\hline
&&&$L/a$&&&\\
\hline
param&6&8&10&12&16&20\\
\hline
\endhead

\hline
$\chi^2$/dof&1.60&1.42&1.30&1.59&1.61&1.10\\
$N_{dof}$&54&55&36&17&7&2\\
\hline
\endfoot
$c_{1,L/a}$&0.380(11)&0.4092(66)&0.4269(97)&0.413(10)&0.467(20)&0.463(32)\\
$c_{2,L/a}$&-0.08(13)&-0.192(46)&-0.224(70)&-0.167(84)&-0.154(46)&-0.111(70)\\
$c_{3,L/a}$&0.56(54)&0.73(11)&0.75(17)&0.82(22)&0.164(28)&0.129(38)\\
$c_{4,L/a}$&-1.2(1.1)&-0.837(98)&-0.81(17)&-1.02(23)&0&0\\
$c_{5,L/a}$&1.6(1.1)&0.342(31)&0.319(54)&0.417(80)&0&0\\
$c_{6,L/a}$&-1.10(57)&0&0&0&0&0\\
$c_{7,L/a}$&0.32(12)&0&0&0&0&0\\
\end{longtable}
\end{center}

\pagebreak

\bibliography{ConfWindPRD}

\end{document}